\documentclass{cernyrep}
\usepackage{texnames}
\usepackage[T1]{fontenc}
\usepackage{varwidth}
\usepackage{xcolor}

\begin{document}

\newcommand {\lumi}     {{\cal L}}
\newcommand {\drom}     {{\mathrm{d}}}
\newcommand {\ilumi}    {{\cal L}_{\mathrm{int}}}
\newcommand {\Ucmsq}     {\mathrm{cm}^2}
\newcommand {\Ucmsqm}     {\mathrm{cm}^{-2}}
\newcommand {\Ulumi}    {\mathrm{cm}^{-2}\mathrm{s}^{-1}}
\newcommand{\epem} {\mathrm{e}^{\mathrm{+}}\mathrm{e}^{\mathrm{-}}}
\newcommand{\heff}{h_{\mathrm{eff}}}
\newcommand{\heffx}{h_{\mathrm{eff},x}}
\newcommand{\heffy}{h_{\mathrm{eff},y}}
\newcommand{\rel}{R_{\mathrm{el}}}
\newcommand{\rinel}{R_{\mathrm{inel}}}

\newcommand{\picturefolder}{./epsfig}
\newcommand{\figurefolder}{./figures}

\title{Machine Protection and Operation for LHC}

\author{J. Wenninger}

\institute{CERN, Geneva, Switzerland}

\maketitle % this produces the title block

\begin{abstract}
Since 2010 the Large Hadron Collider (LHC) is the accelerator with the highest stored energy per beam, with a record of 140~MJ at a beam energy of 4~TeV, almost a factor of 50 higher than other accelerators. With such a high stored energy, machine protection aspects set the boundary conditions for operation during all phases of the machine cycle. Only the low-intensity commissioning beams can be considered as relatively safe. This document discusses the interplay of machine operation and machine protection at the LHC, from commissioning to regular operation.
\end{abstract}

\section{Introduction}

\begin{figure}[btp]
  \begin{center}
\includegraphics[width=0.55\linewidth]{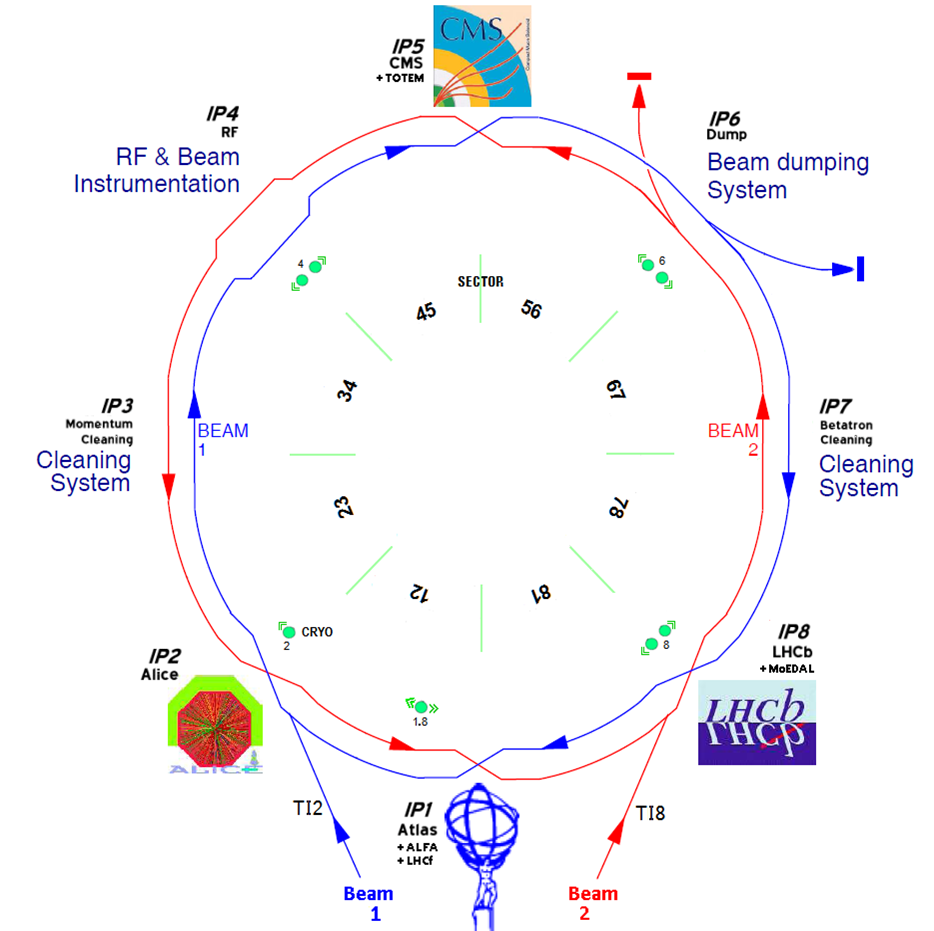}
   \end{center}
\caption{Layout of the LHC with the eight interaction points labelled IP1 to IP8. The experiments ATLAS, ALICE, CMS and LHCb are installed in IP1, IP2, IP5 and IP8, respectively. Beam~1 is injected close to IP2 and circulates clockwise, Beam~2 is injected close to IP8 and circulates counter-clockwise. The two beams exchange position between outside and inside of the ring at every experiment to ensure that the path length is the same for both beams. The two beam dumps are located around IP6.}
\label{fig:layout}
\end{figure}

The Large Hadron Collider (LHC) is the last in the series of hadron colliders after the ISR (Intersecting Storage Ring), SPS (Super Proton Synchrotron), Tevatron, HERA and RHIC (Relativistic Heavy Ion Collider). The machine elements are installed on average 100~m below the surface in the 26.7~km long accelerator tunnel that housed the Large Electron Positron collider (LEP) between 1989 and 2000~\cite{LHCLYN,LHCDR}. The ring consists of eight arcs and of eight long straight sections (LSSs). The large particle physics experiments ALICE, ATLAS, CMS and LHCb are installed at interaction points (IPs) in the middle of four LSSs, while the other LSSs house the collimation (or beam cleaning) system, the radio-frequency (RF) system, the beam instrumentation and the beam dumping system. The layout of the LHC is shown in Fig.~\ref{fig:layout}.

\begin{figure}[btp]
  \begin{center}
\includegraphics[width=0.8\linewidth]{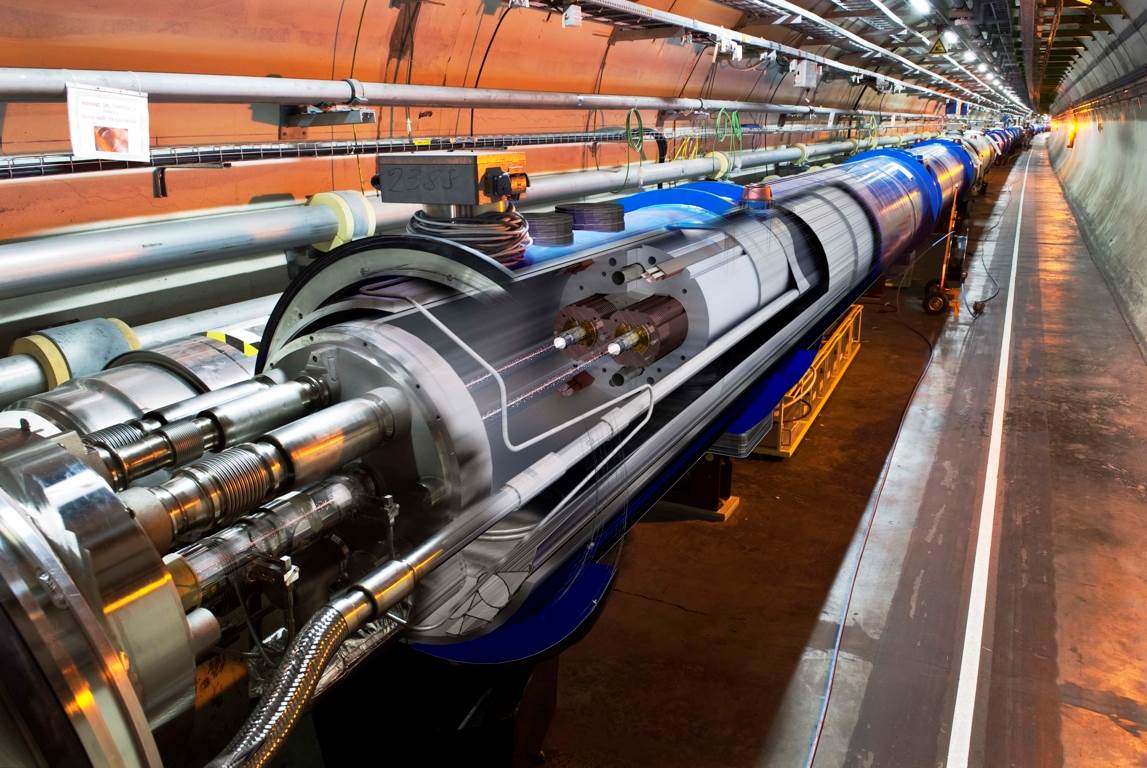}
   \end{center}
\caption{View of the LHC tunnel with an artistic cut through a LHC dipole magnet, highlighting the twin magnet coils as well as the two vacuum chambers.}
\label{fig:magnet-cut}
\end{figure}

 A dipole field of 8.3~T is required to bend hadrons with a momentum of 7~TeV/$c$ per unit charge in the tunnel; this is 60\% higher than in previous accelerators. Such a magnetic field strength is achieved with superconducting dipole magnets made of NbTi. With a 2-in-1 magnet design the two rings fit inside the 3.8~m diameter LEP tunnel; see Fig.~\ref{fig:magnet-cut}. Both rings are accommodated in a single cryostat and the distance between the two vacuum chambers is only 19~cm. The two proton or ion beams circulate in opposite directions in two independent vacuum chambers. Each dipole magnet is 14.3~m long; the associated cryostat is 15~m long. Besides the 1232 dipole magnets that constitute around 85\% of each arc, the magnet lattice also includes quadrupole magnets that focus the beam, sextupole magnets to correct chromatic effects and octupoles to stabilize the beam. A total of 8000 superconducting magnets are used to control the two beams.

Eight continuous cryostats with a length of 2.7~km each cool the superconducting magnets to their operating temperature of 1.9~K. After cool down the LHC cryostats contain 130 tons of liquid helium and around 37\,000 tons of material are cooled to that temperature. The magnets and the cooling system based on superfluid helium form by far the longest refrigerators on Earth. The cryogenic system has to be extremely reliable; in 2012 the system achieved an overall uptime of 95\%.

Around 1600 power converters provide current to the magnets; for the main circuits the peak currents reach 13~kA. The magnetic energy stored in each arc cryostat is around 1~GJ. This energy has to be safely extracted in case one of the magnets quenches, i.e. performs a transition from the superconducting to the normal-conducting state~\cite{MAG-PROT}. Large dump resistors that are capable of absorbing the energy are automatically switched into the main circuits in case of a quench.

\begin{figure}[btp]
  \begin{center}
\includegraphics[width=0.9\linewidth]{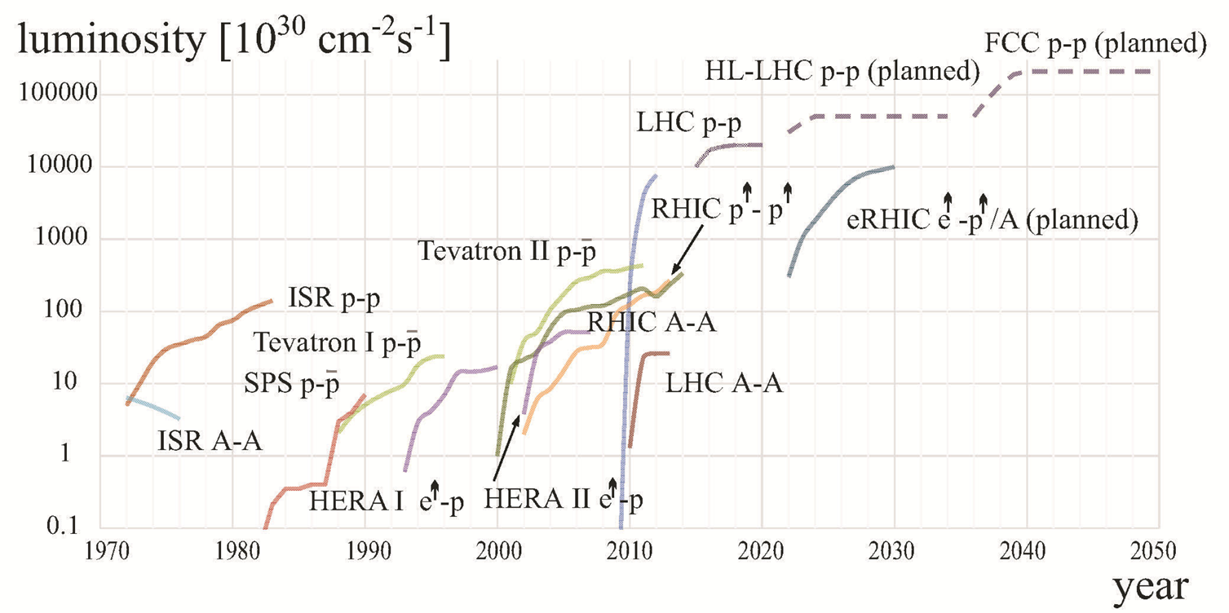}
   \end{center}
\caption{Evolution of the peak luminosity of hadron colliders since 1970 (image courtesy of W. Fischer, BNL). Future projects like the high-luminosity LHC (HL-LHC)~\cite{HL-LHCa,HL-LHCb}, eRHIC and a possible 100~km circumference collider (FCC)~\cite{FCC} are also indicated.}
\label{fig:lumicolliders}
\end{figure}

The performance of a collider is characterized by its luminosity $\lumi$. The event rate of a physical process with cross-section $\sigma$ (with unit of an area [m$^2$]) is given by $\sigma \times \lumi$. The luminosity may be expressed as
\begin{equation}\label{eq:lumi}
  \lumi = \frac{k f N^2}{4 \pi \sigma_x \sigma_y},
\end{equation}
where $f$ is the revolution frequency (11.24~kHz for the LHC), $k$ is the number of bunches, $N$ is the number of particles per bunch and $\sigma_x$ and $\sigma_y$ are the horizontal and vertical beam sizes at the collision point. The highest luminosity is achieved with the smallest possible beam cross-section, a large number of bunches and a high bunch population. Up to 2808 bunches can be filled and accelerated in each LHC beam; the minimum distance between bunches is 25~ns or 7.5~m. Each proton bunch consists of up to $1.7 \times 10^{11}$ protons. The bunches have a typical length of 7 to 10~cm. At the interaction point the transverse rms sizes of the colliding beams are around 20~$\mu$m~\cite{LHC-RUN1}. Figure~\ref{fig:lumicolliders} presents the evolution of hadron collider luminosity over time. The LHC pushes the energy frontier by a factor of seven and the luminosity frontier by a factor of 25; the luminosity gain is mainly obtained with very high beam intensities.

The LHC dipole magnets were produced by three industrial firms and the last dipole magnet was delivered to CERN in November 2006. Each magnet was trained on CERN test benches to a magnetic field of 8.7~T, approximately 5\% above the design target. A few training quenches were typically required to reach the nominal field of 8.3~T. Training quenches are due to the release of extremely small amounts of frictional energy (10--100~nJ) due to coil movements when the magnetic field is increased. In June 2007 the first arc (sector) of the LHC was cooled down and ready for commissioning and in April 2008 the last dipole magnet was lowered into the LHC tunnel. One of the essential components of the commissioning phase was the testing of the LHC superconducting magnets and the associated powering and protection equipment. In early 2008 it became apparent that the LHC dipole magnets had to be re-trained to their nominal field; the first magnet quenches appeared at fields corresponding to beam energies of around 5.5~TeV. A training campaign on one arc revealed that the number of required re-training quenches increased rapidly with the magnetic field. The estimated number of quenches required to reach 6.5~TeV is around 140, confirmed during the re-commissioning in 2015, while for 7~TeV the expected number can be as high as 1000. Since such a training campaign would have required a long time, it was decided to lower the energy for the commissioning and first operation phase to 5~TeV~\cite{EPAC08}.

 On 10 September 2008 beams were circulating for the first time in both LHC rings. The startup was however brought to an abrupt halt on 19 September 2008 when a defective high-current soldering between two magnets triggered an electric arc that released around 600~MJ of stored magnetic energy~\cite{PAC99}. The accelerator was damaged over a distance of around 700~m; 53 magnets had to be replaced or repaired. The beam vacuum was polluted with dust and soot over 2~km. The repair, improvement and re-commissioning of the LHC lasted until November 2009. More details on the incident that did not involve beams are presented in the Appendix. During the repair a systematic soldering problem affecting roughly 15\% of all the high-current (13~kA) cable joints was discovered. Since this problem could not be solved immediately, the beam energy had to be limited to 3.5~TeV until a complete repair campaign could be performed. The LHC therefore operated in 2010 and 2011 at this energy, before the energy was increased to 4~TeV in 2012 and 2013 because no magnet was quenched at 3.5~TeV during beam operation, thus lowering the risk of problems with the cable joints. The energy will be pushed to 6.5~TeV and above in 2015 after the repair campaign of 2013 and 2014~\cite{LS1-IPAC13,SPLICE-IPAC13}.

\begin{figure}[btp]
  \begin{center}
\includegraphics[width=0.95\linewidth]{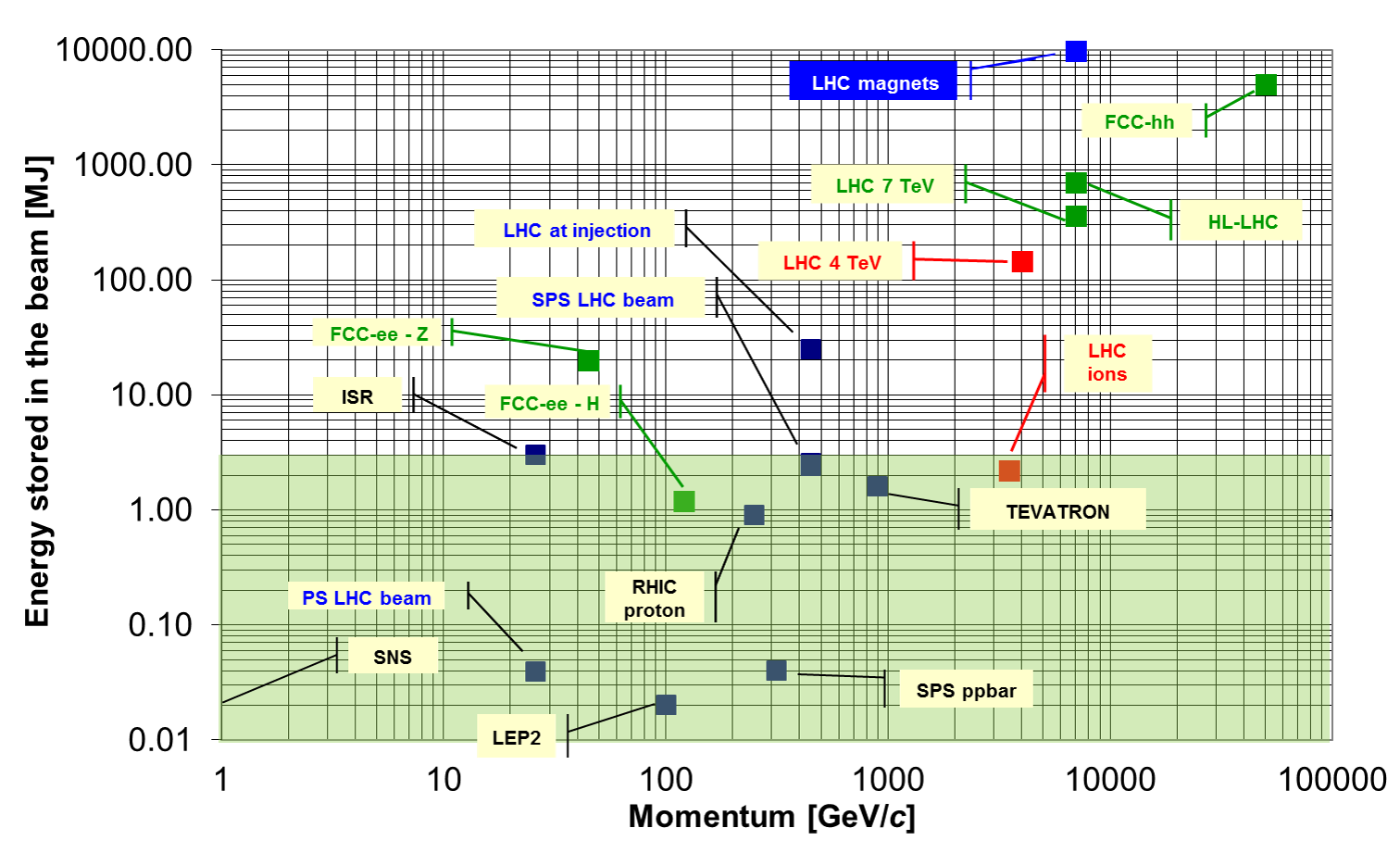}
   \end{center}
\caption{Stored energy of the beams as a function of the momentum for various accelerators. The area shaded in light green corresponds to the state of the art before the startup of the LHC. All accelerators operate with hadrons except the lepton colliders LEP2 and FCC-ee (the CERN 100~km collider study).}
\label{fig:storedE}
\end{figure}

\section{Machine protection at the LHC}

\begin{figure}[btp]
  \begin{center}
\includegraphics[width=0.8\linewidth]{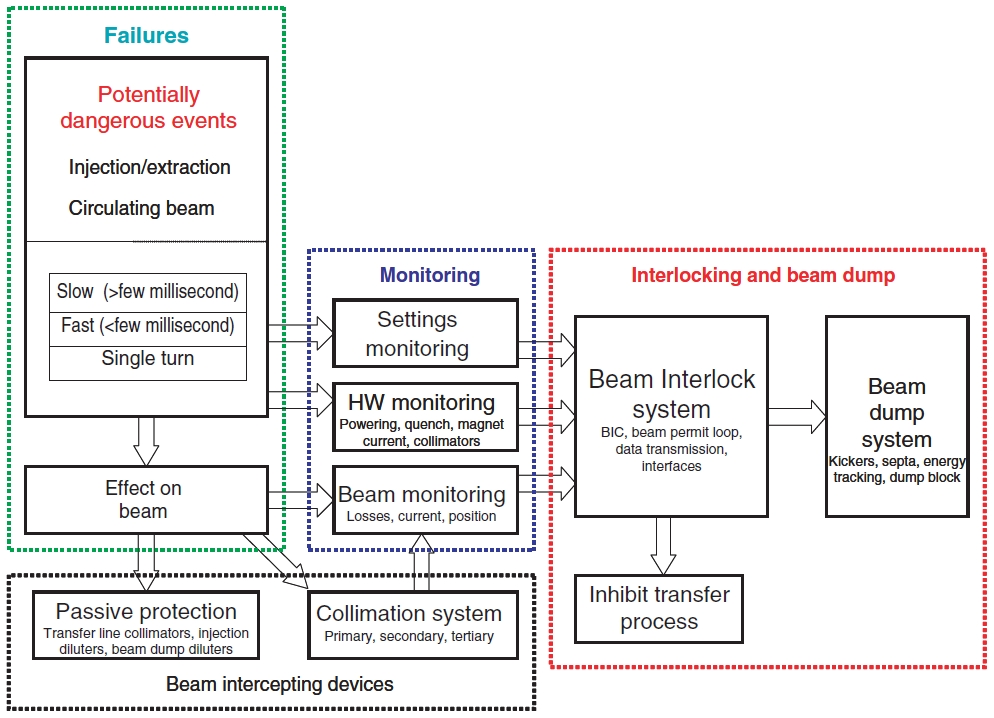}
   \end{center}
\caption{Schematic outline of the main functions of the LHC MPS. Protection against failures is achieved by active monitoring followed by a dump action or by passive protection with absorbers.}
\label{fig:mps-function}
\end{figure}

At 7~TeV each nominal LHC beam stores an energy of 360~MJ. This corresponds to the energy content of 80~kg of explosives and is a hundred times higher than previously achieved accelerator records. The energy stored in the beams of various accelerators is shown in Fig.~\ref{fig:storedE}. Machine protection systems (MPSs) with unprecedented safety levels are therefore required to operate the LHC~\cite{LHCMPa,LHCMPb}. This is achieved with a combination of active protection by equipment and beam parameter monitoring, as well as with passive protection by a large number of collimators as indicated in Fig.~\ref{fig:mps-function}.

\begin{figure}[btp]
  \begin{center}
\includegraphics[width=0.65\linewidth]{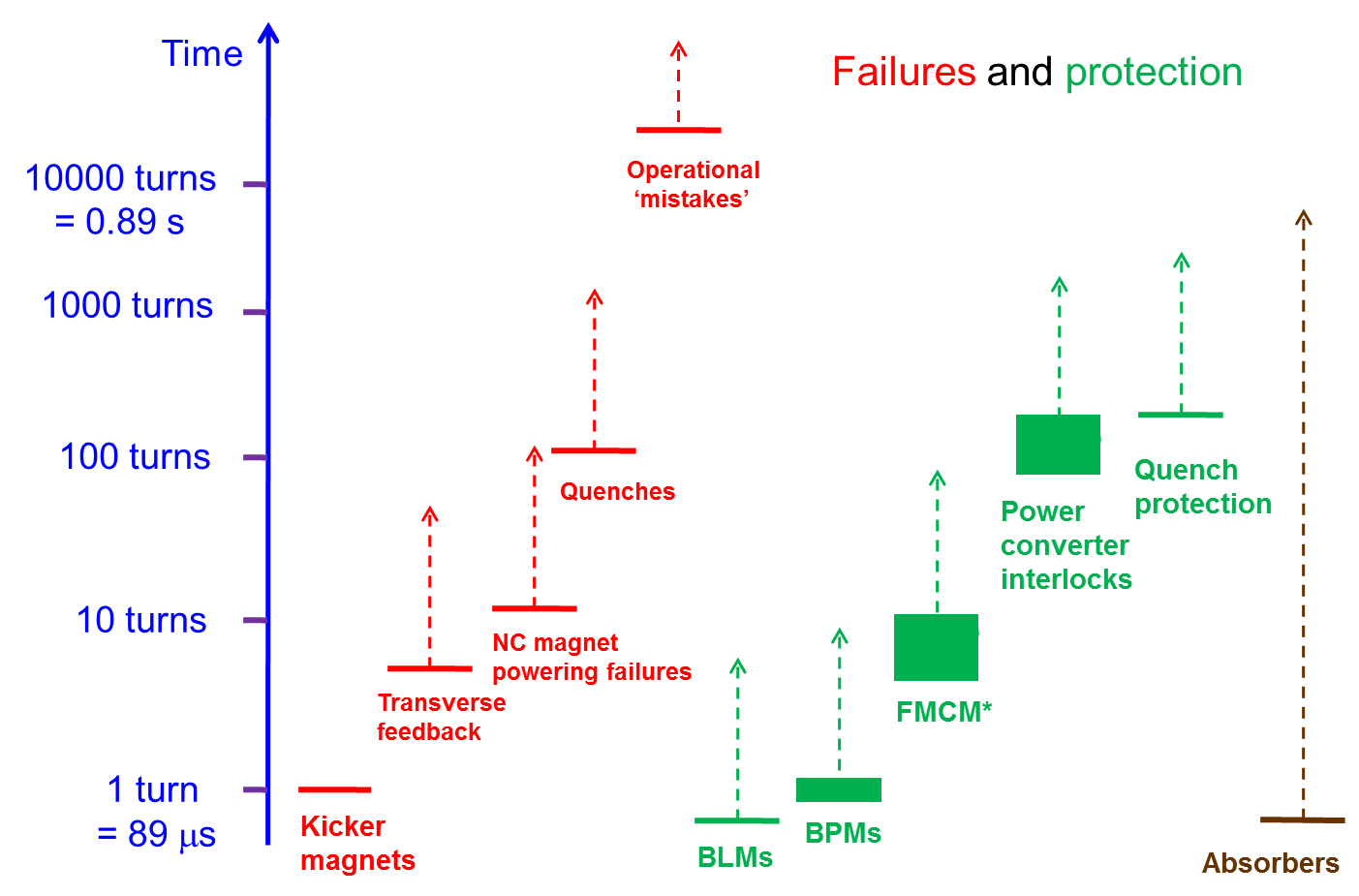}
   \end{center}
\caption{Time-scales for failures (red) and for the reaction of the active protection systems (green) at the LHC. The symbol FMCM stands for fast magnet current change monitor, a device to detect fast current changes in normal-conduction magnet circuits with very short time constants~\cite{FMCM}. The symbol NC stands for normal-conducting magnet.}
\label{fig:mps-timescale}
\end{figure}

The function of the LHC MPS is to protect the LHC accelerator, its injection and extraction transfer lines, as well as its experiments~\cite{MPS-EXP} against a large spectrum of failures~\cite{JAS-KAIN-RING}. Failures may be detected by monitoring equipment or beam parameters, for example power converter currents and state or beam loss rates. In parallel their impact is mitigated by passive absorbers in the form of collimators and absorbers. In the case where abnormal conditions are detected, interlock signals are sent to the the beam interlock system (BIS)~\cite{LHCMPa,LHCMPb} that will trigger a beam dump or stop injection of the beams. A summary of the time-scale of failures and of the reaction time of MPS components is shown in Fig.~\ref{fig:mps-timescale}.

The LHC BIS interfaces the sources of interlocks (clients) and the beam dumping and injection kicker systems. There are currently 189 inputs from client systems. The list of connected systems includes:
\begin{itemize}
  \item the powering interlocks for normal-conducting and superconducting circuits;
  \item the fast current change detection of electrical circuits with fast failure time constants;
  \item the electrical circuits of the main experiment magnets;
  \item the interlock signals from the experiments (excessive rates);
  \item approximately 4000 beam loss monitors (BLMs)~\cite{JAS-DEHNING};
  \item the beam position monitors (BPMs);
  \item the beam screens;
  \item the collimators and absorbers (positions and temperatures);
  \item the movable experimental detectors;
  \item the RF system;
  \item the beam dumping system (state and trigger unit);
  \item the injection and aperture kicker systems;
  \item the personnel access system and its associated beam stoppers;
  \item the operator inhibit buttons;
  \item the vacuum valves;
  \item the programmable beam dump.
\end{itemize}

\begin{figure}[btp]
  \begin{center}
\includegraphics[width=0.7\linewidth]{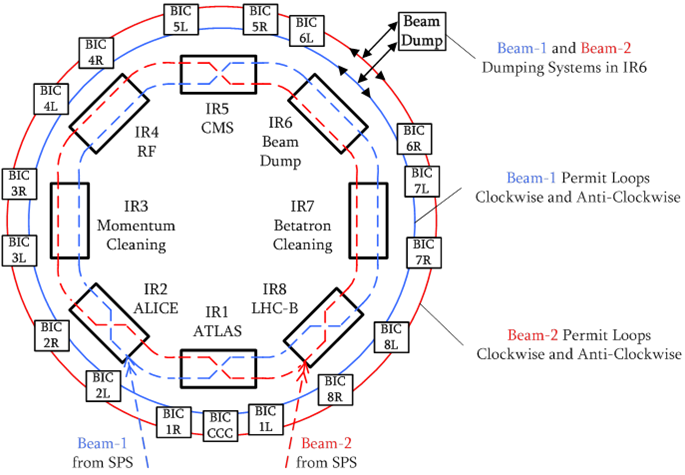}
   \end{center}
\caption{Schematic layout of the LHC beam interlock system with 17 beam interlock controllers (BICs) distributed around the ring (image courtesy of B. Todd, CERN). The BIC modules are connected to beam permit loops that interface with the LHC beam dumping system.}
\label{fig:bisloop}
\end{figure}

   Beam interlock controller (BIC) modules are installed in each LHC access point and provide the interface between the client systems and the BIS; see Fig.~\ref{fig:bisloop}. The BIC units are connected to each other and to the beam dumping system though permit loops. There are two permit loops per beam; the signals in the loops propagate clockwise in one loop and counter-clockwise in the other loop. This arrangement ensures always the fastest possible transmission of a dump request to the dumping system. At the LHC the dump delay can reach around three turns or 270~$\mu$s; this delay includes propagation of an interlock on the permit loops, synchronization of the beam dump with the beam abort gap (see below) and one turn to extract the entire beam.

The permit loops are connected directly to the LHC injection interlock system, which is made from the same building blocks (BIC modules). The LHC injection interlocks act on the LHC injection kicker and on the SPS extraction interlock system. This ensures that no beam may be extracted from the SPS injector when the beam permit loops are not armed~\cite{JAS-KAIN-INJ}.

\begin{figure}[btp]
  \begin{center}
\includegraphics[width=1\linewidth]{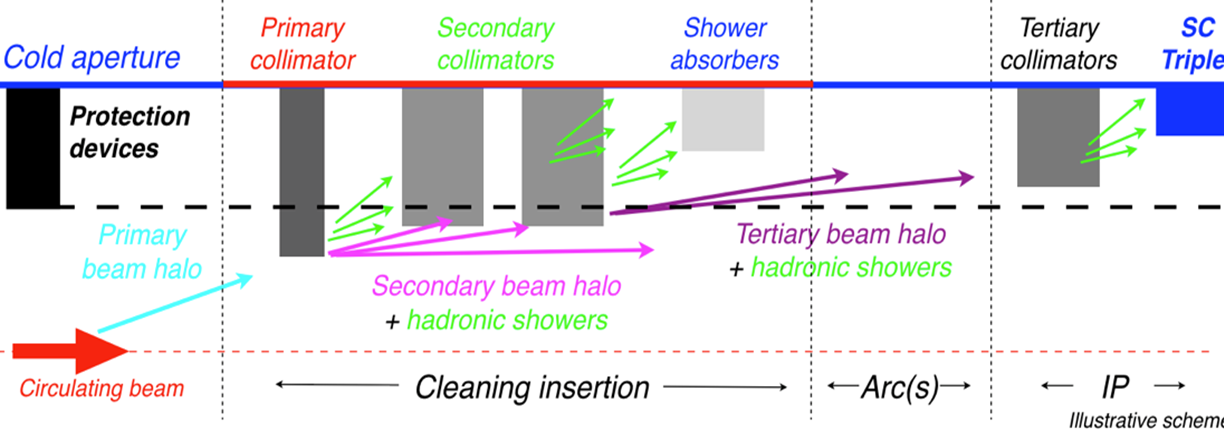}
   \end{center}
\caption{Principle of multistage beam collimation (cleaning) at the LHC}
\label{fig:cleaning}
\end{figure}

The superconducting magnets in the LHC lose their superconducting state (quench) when they are heated with about 10 mJ per cubic centimetre. This is to be compared with the stored energy of many 100~MJ. To prevent unavoidable particle losses from the beam hitting the vacuum chamber within the magnets, possibly leading to sufficiently large energy deposition to trigger a quench~\cite{QUENCH}, a collimation system~\cite{REDAELLI-JAS} must intercept the beam losses with very high efficiency. Contrary to previous hadron colliders that used collimators only for experimental background conditions, the LHC cannot operate without its collimation system. Approximately 100 LHC collimators are installed in two long straight sections as shown in Fig.~\ref{fig:layout}. To be able to absorb the energy of the 7~TeV hadrons, the LHC requires a multistage collimation system  that intercepts the particles in a four-step process as outlined in Fig.~\ref{fig:cleaning}, covering the entire six-dimensional phase space, catching particles with large transverse oscillation amplitudes and energy errors~\cite{REDAELLI-JAS}.
Each collimator is composed of two parallel, typically 1.2~m long material blocks (jaws) that define a variable gap through which the beams circulate. The smallest gap widths are roughly 2~mm. Depending on their location in the ring and on their role, the collimator jaws are made of fibre-reinforced carbon, strongly absorbing copper and tungsten blocks. The system worked perfectly so far, also thanks to excellent beam stability and machine reproducibility. Around 99.99\% of the protons that were lost from the beam were intercepted. A single set-up of the collimation system was required per year. From the machine protection perspective, the LHC collimator must fulfil a dual role, halo collimation (beam cleaning) and passive protection of the accelerator component.

\begin{figure}[btp]
  \begin{center}
\includegraphics[width=0.9\linewidth]{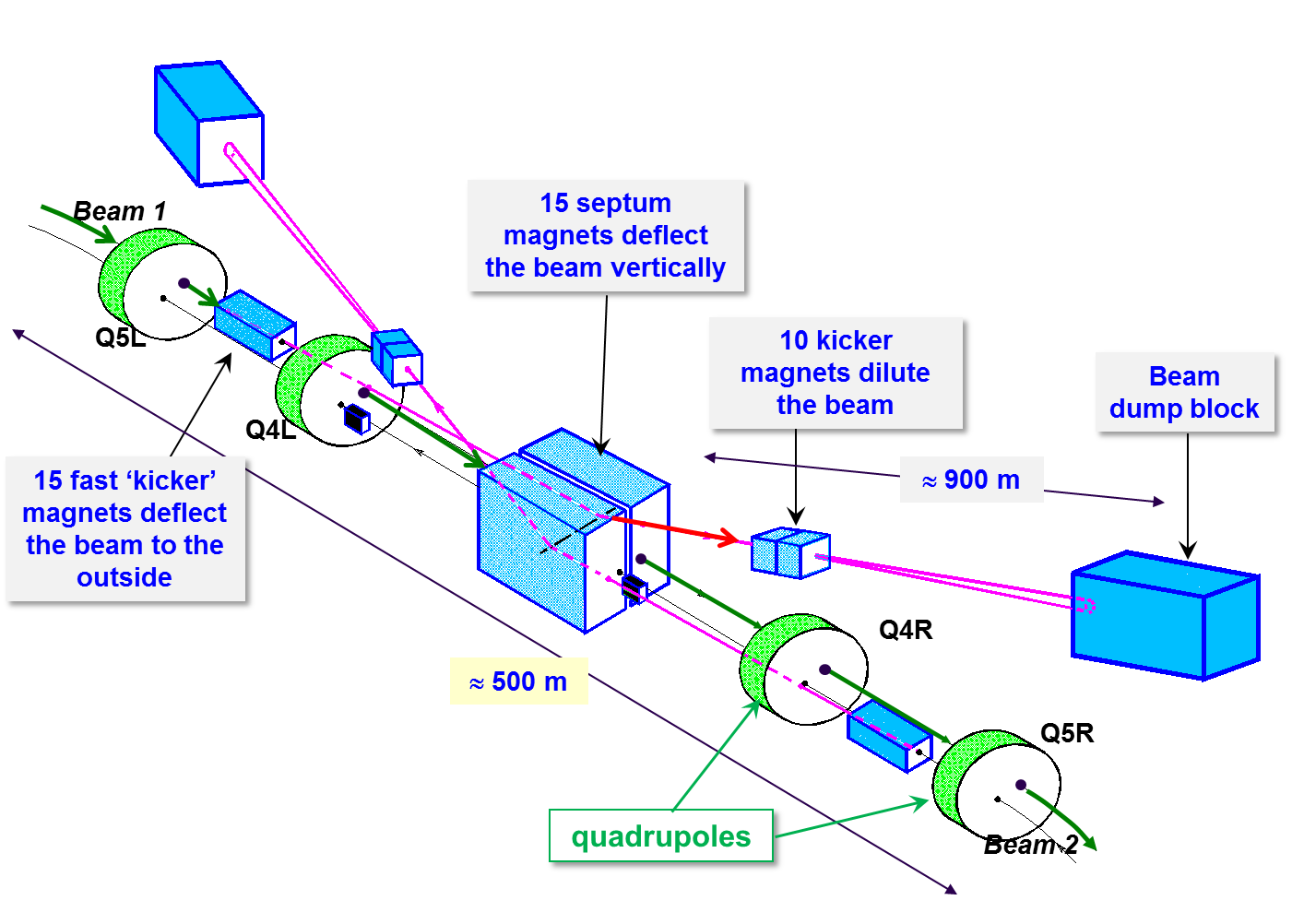}
   \end{center}
\caption{Layout of the LHC beam dumping system (LBDS). For each beam 15 fast kicker magnets deflect the beam out of the ring into septum magnets that deflect the beam further into the vertical plane. On their way to the beam dump, fast dilution kickers spread out the bunches on the surface of the dump block (courtesy of M.~Gyr).}
\label{fig:lbds-layout}
\end{figure}

The LHC beam dumping system (LBDS) is a complex system composed of 15 fast kicker magnets that deflect the beam to the outside of the ring, 15 normal-conducting septum magnets that deflect the beam vertically out of the plane of the ring, 10 dilution kicker magnets that spread the beam over the beam dump surface and finally the beam dump block~\cite{LHCLYN, LHCDR, LHCMPa,LHCMPb}. The layout of the LBDS is shown in Fig.~\ref{fig:lbds-layout}. The 10~m long carbon dump block is the only LHC element capable of absorbing the nominal beam. The beam is swept over the dump surface to lower the power density. Without the sweep the beam could drill a hole with a depth of a few metres into the dump block through hydro-dynamic tunnelling~\cite{MATDAM}.

A 3~$\mu$s long particle-free gap in the beam (beam abort gap) provides a time window for the LBDS dump kickers to raise to their nominal field. The dump kickers must be accurately synchronized to the beam abort gap to avoid spreading beam across the aperture during the kicker rise time. Possible failure modes are:
\begin{itemize}
  \item the abort gap fills with beams (RF fault, debunching, injection error);
  \item the kicker synchronization fails or a kicker fires spontaneously (not synchronized) -- so-called \textit{asynchronous beam dumps}.
\end{itemize}
The asynchronous dump is the \textit{ultimate unavoidable failure}: the LHC must be protected from this failure \textit{passively} by absorbers. Two large absorbers in front of the extraction septum and in front of the first superconducting magnet downstream of the dump kickers and septa protect the LHC against damage from asynchronous dumps and from residual beam in the abort gap.
Dump kicker powering, synchronization and triggering are designed to exclude out-of-sync triggers with high reliability.
 The spontaneous trigger is estimated to occur roughly once per year. So far none has been observed during high-intensity operation, but the system operated at reduced high voltage (4~TeV instead of 7~TeV), which also lowers the probability of spontaneous kicker firing.

A direct link between the LBDS and the injection system ensures that the LHC injection kicker magnets cannot be triggered with a timing that would inject beam into the abort gap. The beam population in the abort gap is monitored using synchrotron light. When the particle density in the abort gap exceeds a pre-defined threshold, the transverse feedback system of the LHC is used to excite those particles until they hit the collimators.

\section{Interlock masks for commissioning and operation}

Already during the design phase of the LHC MPS, the need for masking interlocks was recognized~\cite{LHCMPa,LHCMPb}. Some flexibility is always required for low-intensity commissioning and setting up phases where the risk of damage is also much lower.
To avoid masking interlocks by raising thresholds and opening tolerance windows for many parameters (with the risk of errors during the reversal), the concept of masking interlocks when \textit{the beam is safe} was introduced at an early stage. Such a safe beam should not be able to damage accelerator components. The corresponding beam intensity limit depends on the beam energy and on the beam emittance. It also depends on the material that is considered for damage. At the LHC, copper was selected as reference material. Given the large difference between the energy required to quench a superconducting magnet and the energy required to damage the same magnet, it is clear that even a safe beam will be able to quench one or more magnets.

In the early design phase of the LHC machine protection system, around the years 2000--2004, not much was known about equipment damage by high-energy and high-intensity beams. In order to benchmark damage by a high-energy LHC-type beam, a dedicated experiment was set up in a SPS transfer line~\cite{PAC05}. In this controlled experiment beams of 450~GeV protons with LHC bunch structure (25~ns spacing) and nominal LHC beam emittances were directed into a special target composed of a sandwich of materials (copper, stainless steel and zinc). From this experiment the damage limit for copper was established at $2 \times 10^{12}$~protons. It should be noted that the damage limit for stainless steel was more than a factor of four higher. The observations were in good agreement with rather simple estimates for the damage limit based on shower simulations to define the deposited energy and heating of the materials up to the melting point.

\begin{figure}[btp]
  \begin{center}
\includegraphics[width=0.7\linewidth]{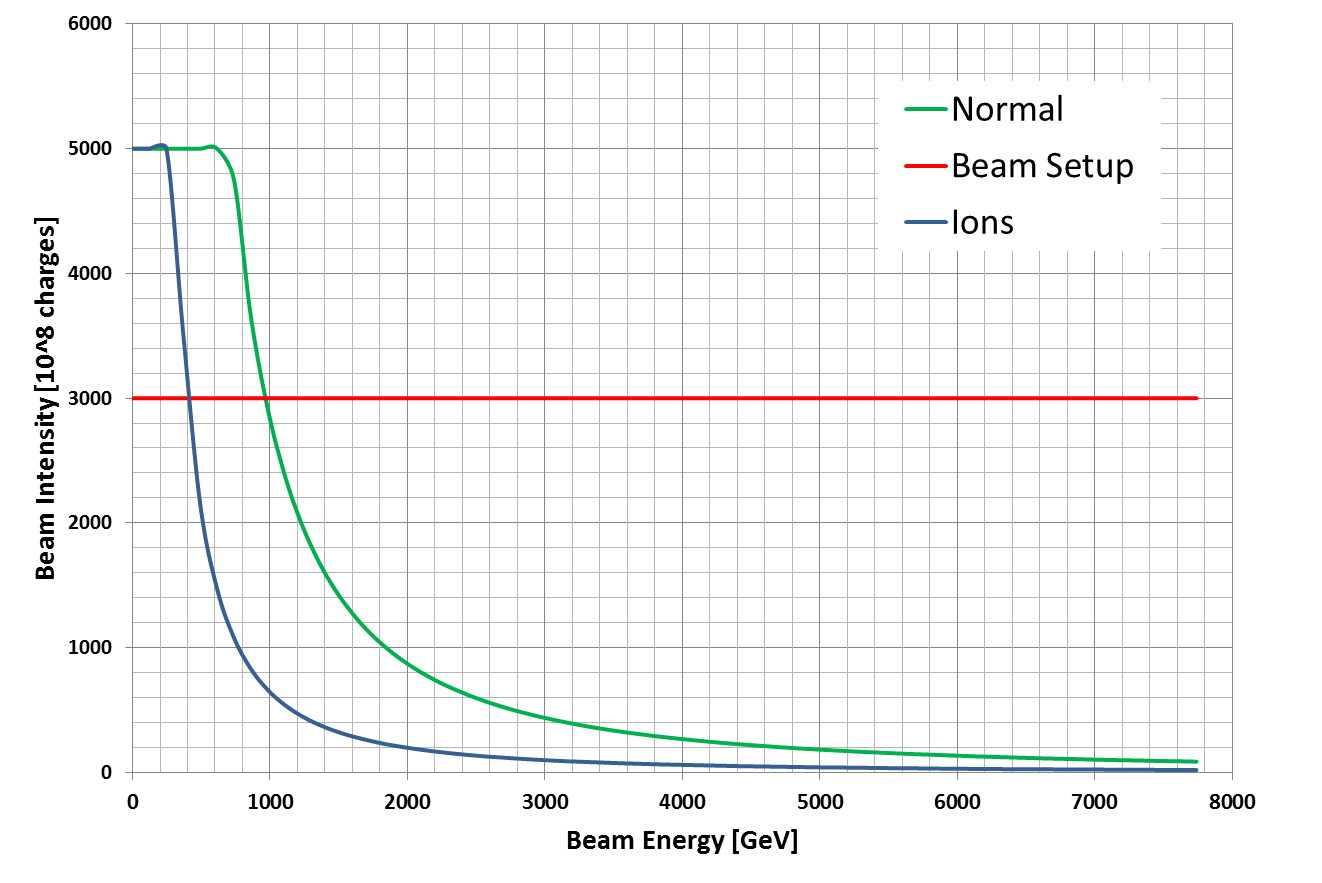}
   \end{center}
\caption{Limiting beam intensity (in $10^{8}$ charges) as a function of energy for the set-up beam flag. The \textit{Normal} SBF corresponds to the scaling of Eq.~(\ref{eq:sbf-scaling}). The \textit{beam set-up} equation is a relaxed version used for beam commissioning and MPS set-up. For ion beams the limits are scaled by the ratio $Z/A$ of the charge per nucleon.}
\label{fig:sbf-energy}
\end{figure}

FLUKA simulations~\cite{FLUKA} were used to extrapolate the experimental results from 450~GeV to the LHC operating energy of 7~TeV. The simulations predicted the following scaling law for the safe beam intensity $I_{\rm SB}$ with beam energy $E$:
\begin{itemize}
  \item larger energy deposition implies a scaling $\propto 1/E$;
  \item shrinking of the transverse emittance (beam area) implies a scaling $\propto 1/E$;
  \item increase of the shower length provides some dilution $\propto \log(E)$.
\end{itemize}
The following effective scaling law with energy (at fixed normalized emittance) was obtained:
\begin{equation}\label{eq:sbf-scaling}
  I_{\rm SB} \propto E^{-1.7}.
\end{equation}
This equation is implemented inside the LHC safe machine system (SMP) and is indicated as the \textit{Normal} set-up beam flag (SBF) in Fig.~\ref{fig:sbf-energy}. At injection the intensity limit was set initially to $10^{12}$ charges, but was later lowered to $5 \times 10^{11}$ because the beam emittances were roughly a factor of two smaller than expected in the LHC design. The SMP system is connected to reliable Beam Current Transformers (BCTs) and energy sources (based on the dipole fields with four-fold redundancy). It generates the SBF that is distributed over the LHC machine timing system to the BIS. The SBF states are:
\begin{itemize}
  \item true = set-up beam: maskable BIS input signals can be masked;
  \item false = unsafe beam: no BIS input signals may be masked.
\end{itemize}
The beam interlock system is configured to allow masking certain classes of interlocks (maskable) when the SBF is true.

During initial operation it was realized that the definition of the SBF was too restrictive for beam commissioning that required higher bunch charges than expected (nominal bunches with $10^{11}$~protons) due to systematic effects in the LHC beam instrumentation, mainly the beam position monitoring system. As a consequence, relaxed SBF equations were introduced to be able to mask certain interlocks with 2--3 nominal bunches at top energy as shown in Fig.~\ref{fig:sbf-energy}. During commissioning periods a MPS expert can relax the equation of the SBF within the SMP system for certain tests.

\section{Commissioning and organization}

The MPS activities of the LHC were organized since the year 2000 inside a Machine Protection Working Group (later renamed to Machine Protection Panel, MPP). This group steered the design and followed up on implementation, issues and performance of the LHC MPS. All groups involved in MP activities were represented in the MPP.

An executive MP body, the restricted MPP (rMPP), was created when beam commissioning started. Each core MP system has one representative within the rMPP. The rMPP takes decisions related to MPS (example: BLM threshold changes) and steers the intensity ramp up of the machine. Its recommendations are submitted to the CERN management; in almost all cases the recommendations were accepted.

Before the machine startup, procedures were developed for the commissioning of the machine protection subsystems.
The procedures contain test descriptions and frequency of tests (after stop or intervention).
The procedures were then translated into a series of individual tests to be performed on the machine with and without beam,
if required also at different intensity steps.

Machine protection tests are currently documented and tracked on a web page. One MPS expert of the commissioning team checks that all tests required for a given commissioning phase have been performed by the experts. It is in general the system expert that executes the tests for his system and not an independent person. This simple mechanism for tracking the commissioning will be improved in the future. The new concept with test tracking and electronic expert signatures is already in place for the commissioning of the LHC electrical circuits and magnets. The \textit{AccTesting} framework~\cite{ACCTESTING} is based on pre-defined and agreed test sequences. Tests that are ready for execution can be scheduled for execution; test sequences are blocked until tests are analysed and signed by experts. The results are tracked and stored; no test step can be forgotten.
Unfortunately due to lack of resources the test tracking was not yet implemented for the LHC MPS, but it is foreseen in the near future.

Passive protection of the LHC aperture with collimators and absorbers is a key ingredient for operating the LHC safely at high intensity. All failures affecting the machine on a global scale (orbit, optics, emittance, perturbations) must be intercepted by a protection device. The LHC machine setting up involves:
\begin{itemize}
  \item a well-corrected orbit (rms below 0.5~mm);
  \item a well-corrected optics (betatron function beating below 20\% at injection and below 10\% with colliding beams);
  \item a good knowledge of the aperture bottlenecks (after orbit and optics correction).
\end{itemize}
The machine aperture is measured after orbit and optics corrections have been finalized. The global aperture is measured at injection energy (typical aperture limit is at $\approx 12$ $\sigma$, where $\sigma$ is the beam size with a nominal normalized emittance of 3.5~$\mu$m). Local apertures are measured at injection (injection region, beam dump section and around the four experiments) and at top energy with fully squeezed beams (only the four experiments). At top energy the aperture depends on the optics at the different collision points. The aperture limit is generally located in the low-beta quadrupoles next to the collision points.

\begin{figure}[btp]
  \begin{center}
\includegraphics[width=1\linewidth]{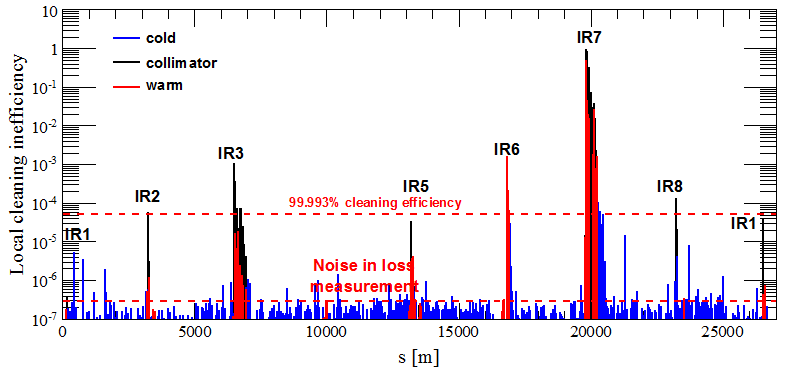}
   \end{center}
\caption{Example of the beam loss distribution during a loss map performed at a beam energy of 4~TeV. The vertical scale gives the losses normalized to the peak loss at the primary collimator. The beam travels from left to right. Blue markers correspond to BLMs located on superconducting elements, red markers to room-temperature elements and black markers to collimators.}
\label{fig:lossmap}
\end{figure}

During machine set-up, all collimators and absorbers are aligned around the closed orbit with appropriate retractions.
For good performance the orbit must be reproducible at the level of 50~$\mu$m. The machine set-up (orbit, optics, aperture and protection devices) is then validated by a campaign of loss maps and simulated asynchronous beam dump tests.
\begin{itemize}
  \item During a \textit{loss map} the transverse beam emittance is blown up until losses are observed on the collimators and absorbers; an example is shown in Fig.~\ref{fig:lossmap}. The loss distribution provides a validation of the collimator alignment and hierarchy~\cite{REDAELLI-JAS}. Initially the emittance was blown up by crossing the third-order resonance in the horizontal or vertical plane. This technique was however not always very reproducible; sometimes the losses were too small, sometimes they were massive and a large fraction of the beam was lost. From 2012 onwards emittance blow up was obtained by noise excitation using the transverse feedback system. This provided fine control over the losses and blow up could be applied to individual bunches~\cite{COLL-MW}.
  \item For a simulated \textit{asynchronous test dump} a low-intensity beam (typically 1--3 bunches) is de-bunched by switching of the RF system. After a few minutes the particles drift into the abort gap, the density of particles in the gap being monitored by a dedicated device based on synchrotron light. When the population in the abort gap is sufficient, a dump is triggered. The beam present in the region of the abort gap mimics the effect of an asynchronous dump. The loss distribution along the ring provides a validation of the dump protection alignment.
\end{itemize}

\section{Intensity ramp}

The entire LHC cycle, from injection to collisions, is always set up with low-intensity beams. The maximum intensity is three bunches with nominal intensity ($\approx 1.1 \times 10^{11}$~protons).
The set-up is made with less than 1~permill of the nominal intensity, which represents a challenge for the beam instrumentation, since there should not be a significant bias of measurements between the set-up and the nominal full-intensity beam (for example for the beam position monitoring).

\begin{figure}[btp]
  \begin{center}
\includegraphics[width=1\linewidth]{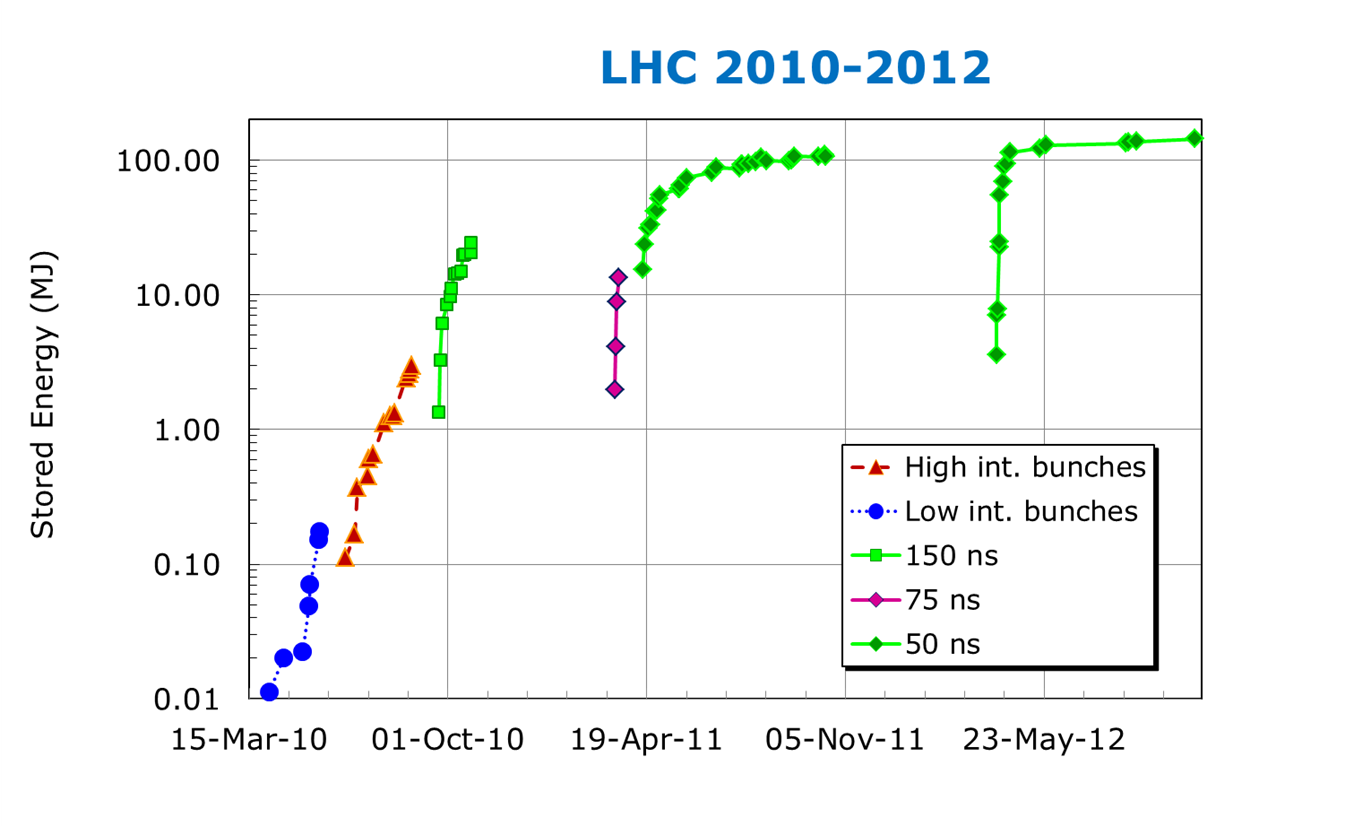}
   \end{center}
\caption{Evolution of the peak stored energy in one LHC beam during Run~1. The different colours detail the beam structure in single bunches (first two periods) or in trains with bunch spacings of 150, 75 and 50~ns.}
\label{fig:storedE-run1}
\end{figure}

\begin{figure}[btp]
  \begin{center}
\includegraphics[width=1\linewidth]{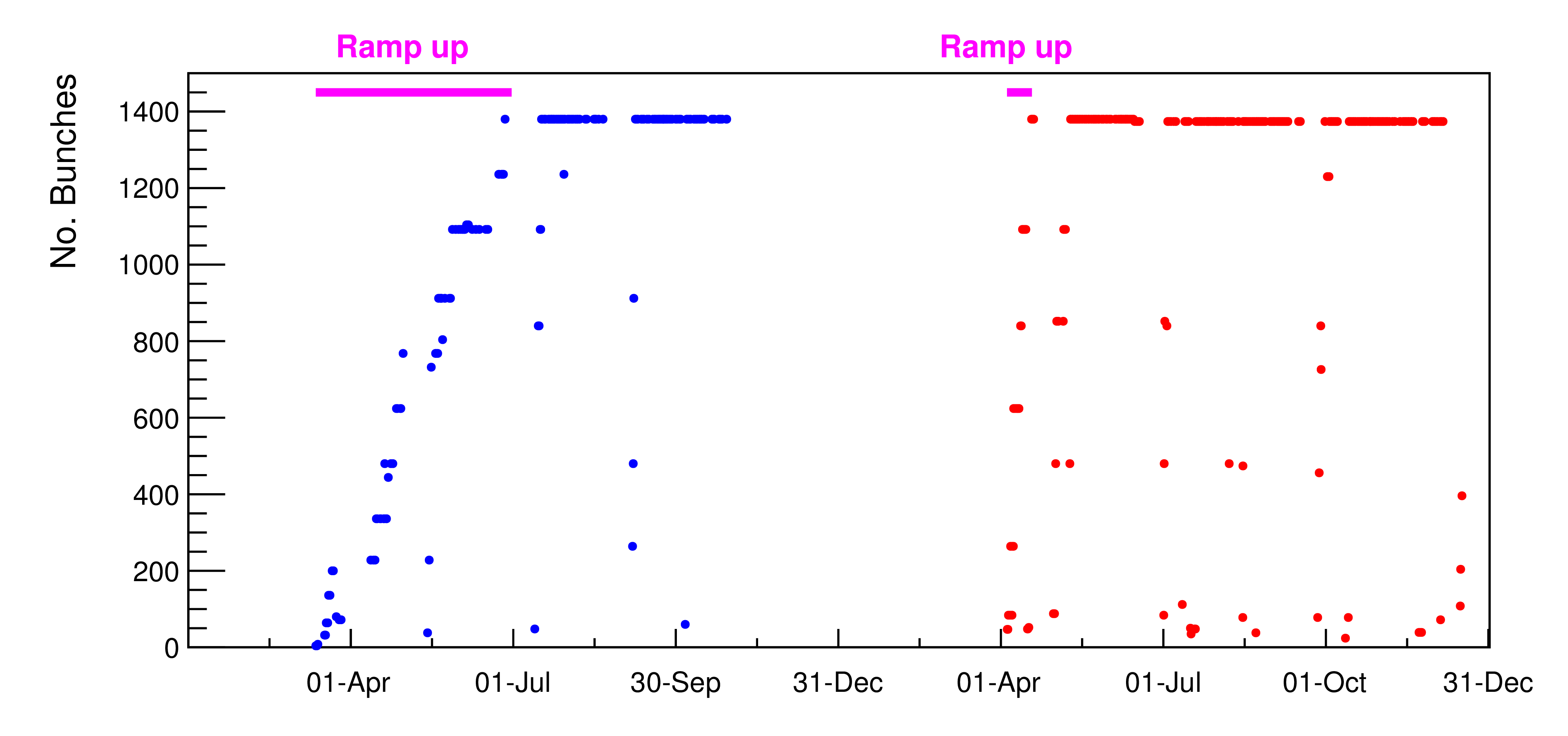}
   \end{center}
\caption{Intensity ramp up of the LHC beams, expressed here through the number of bunches in 2011 and 2012}
\label{fig:rampup-11-12}
\end{figure}

The intensity increase is steered through the restricted Machine Protection Panel (MPPr), which
defines the intensity steps and the requirements to proceed with the intensity increase. The plan for the first learning year in 2010 foresaw three phases:
\begin{itemize}
  \item low-intensity beams for commissioning and early experience. This phase was followed by an internal review of the MPS performance;
  \item an intensity ramp up to a stored energy of 1--2~MJ followed by 4 weeks of operation at that stored energy. Such a stored energy corresponded to state of the art in 2010. This phase was followed by an external review of the MPS performance;
  \item increase of the stored energy into the regime above 10~MJ.
\end{itemize}
Figure~\ref{fig:storedE-run1} shows the evolution of the peak stored energy for one LHC beam between 2010 and 2012. The slow ramp up of 2010 is clearly visible. With the experience that was accumulated in 2010, the stored energy was ramped up above 100~MJ in 2011 over 11 intensity steps, up to a maximum of 1300 bunches. The 2011 intensity ramp up took around 9 effective weeks; the rate was dictated by operational (and not MP) issues as soon as around 600~bunches were stored~\cite{LHC-RUN1}. Losses and the need to adjust of BLM thresholds, vacuum element heating by the beam, beam stability, etc slowed down the pace. Finally, the 2012 intensity ramp up took just 2 weeks with seven intensity steps, as can be seen in Fig.~\ref{fig:rampup-11-12}.

\section{Beam losses at the LHC}

\subsection{Witness beam}

Injection into a synchrotron that has no circulating beam has the same MP issues as a linac~\cite{JAS-KAIN-INJ}. Injection of an intense beam can represent a serious risk or require very important monitoring efforts (all power converters etc) to ensure that the beam is not lost directly on the aperture during the first turn, before the MPS is able to react. To overcome this issue, the concept of \textit{witness} beam (or \textit{beam presence}) was introduced for the LHC~\cite{BPF-CONCEPT}. The principle is that only a probe bunch (typically $10^{10}$~protons, max $10^{11}$~protons) may be injected into an \textit{empty} ring. Such a low intensity cannot provoke damage to components.
A high-intensity injection requires a minimum beam intensity to be circulating; this is the best guarantee that the conditions are reasonable to avoid failures happening on the first turn, before the MPS is able to react.
At the LHC the concept is based on a highly reliable and redundant intensity measurement. A flag indicating the beam presence (true/false) is transmitted to the extraction interlock system of the SPS injector where it is combined with a flag indicating that the SPS beam is has a probe intensity (max $10^{11}$~protons).

Despite storing up to 140~MJ at 4~TeV, not a single superconducting magnet was quenched at the LHC with circulating beam, despite quench thresholds of only a few tens of mJ. On the other hand, many magnets were  quenched at injection, mainly due to (expected) injection kicker failures (seven events in 2012). The beam (roughly 2~MJ) is safely absorbed in injection dump blocks, but the shower leakage can quench magnets over a distance of around 1~km.

\subsection{Beam loss in the cycle}

\begin{figure}[btp]
  \begin{center}
\includegraphics[width=1
\linewidth]{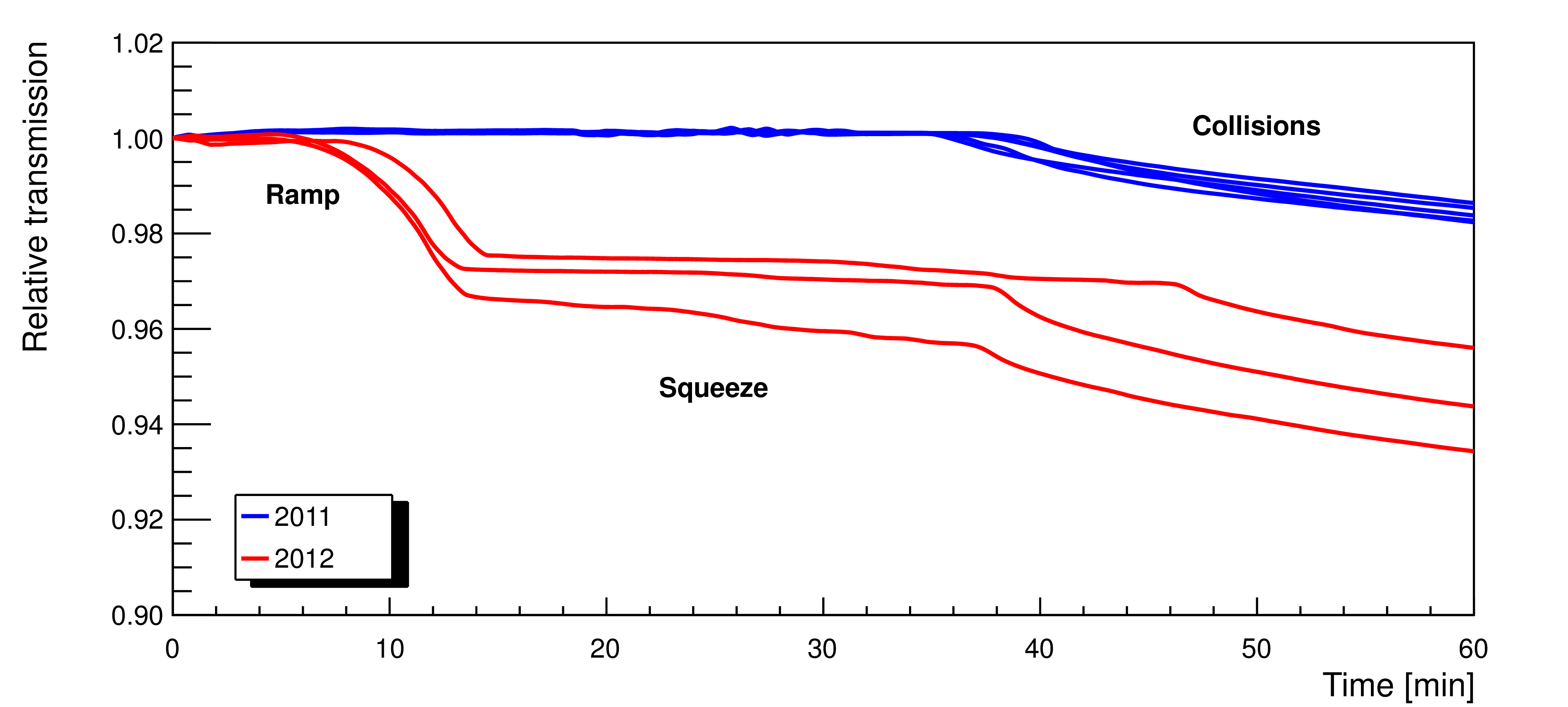}
   \end{center}
\caption{Intensity transmission through the LHC cycle including the energy ramp and the betatron squeeze, the operation phase when the IP betatron functions are reduced to their value in collision.}
\label{fig:transmission}
\end{figure}

\begin{figure}[btp]
  \begin{center}
\includegraphics[width=0.8\linewidth]{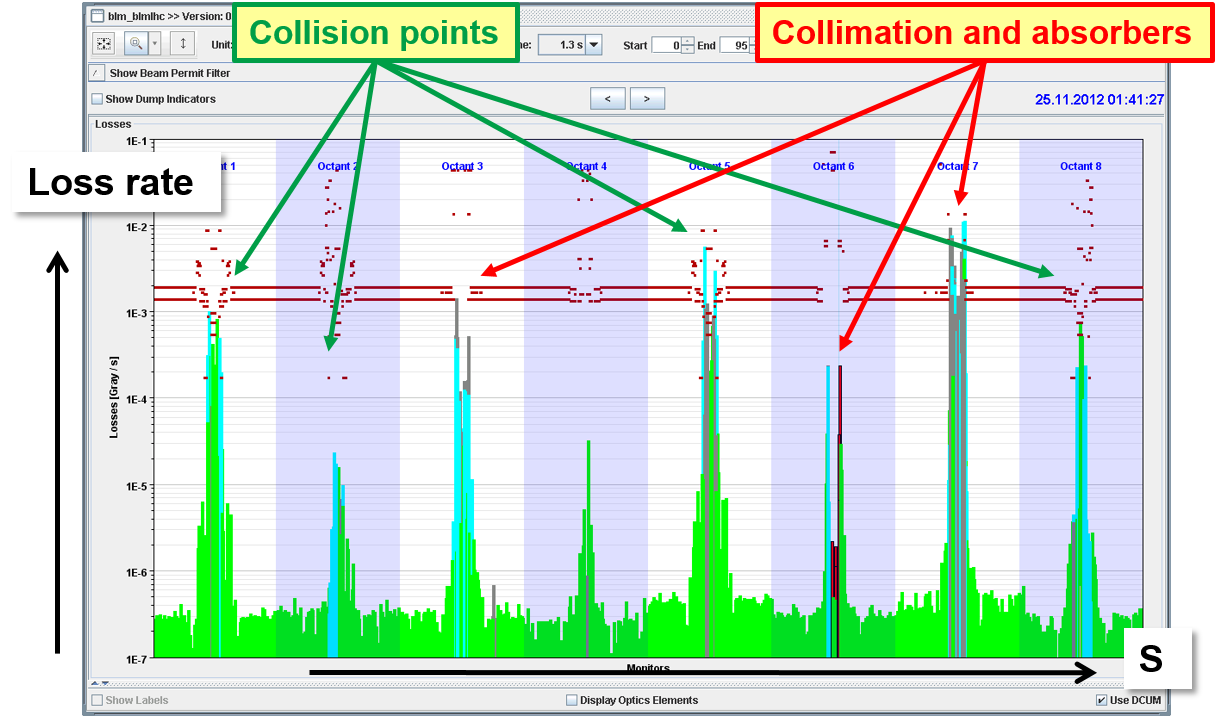}
   \end{center}
\caption{Distribution of steady-state beam loss around the LHC ring with stable colliding beams}
\label{fig:loss-sb}
\end{figure}

At the LHC characteristic beam losses are observed in the various phases of the machine cycle. Those losses are part of regular operation and must be tolerated, even if one tries to minimize them. The normal losses are:
\begin{itemize}
  \item injection losses (tails of the injected beams, injection oscillations, de-bunched beam);
  \item start of ramp losses (uncaptured beam loss);
  \item scraping on collimators (gap changes, orbit and tune shifts);
  \item losses from the beam halo when beams start to collide (beam--beam effect);
  \item losses due to the beam burn-off that are proportional to luminosity and performance.
\end{itemize}
\begin{figure}[btp]
  \begin{center}
\includegraphics[width=1.0\linewidth]{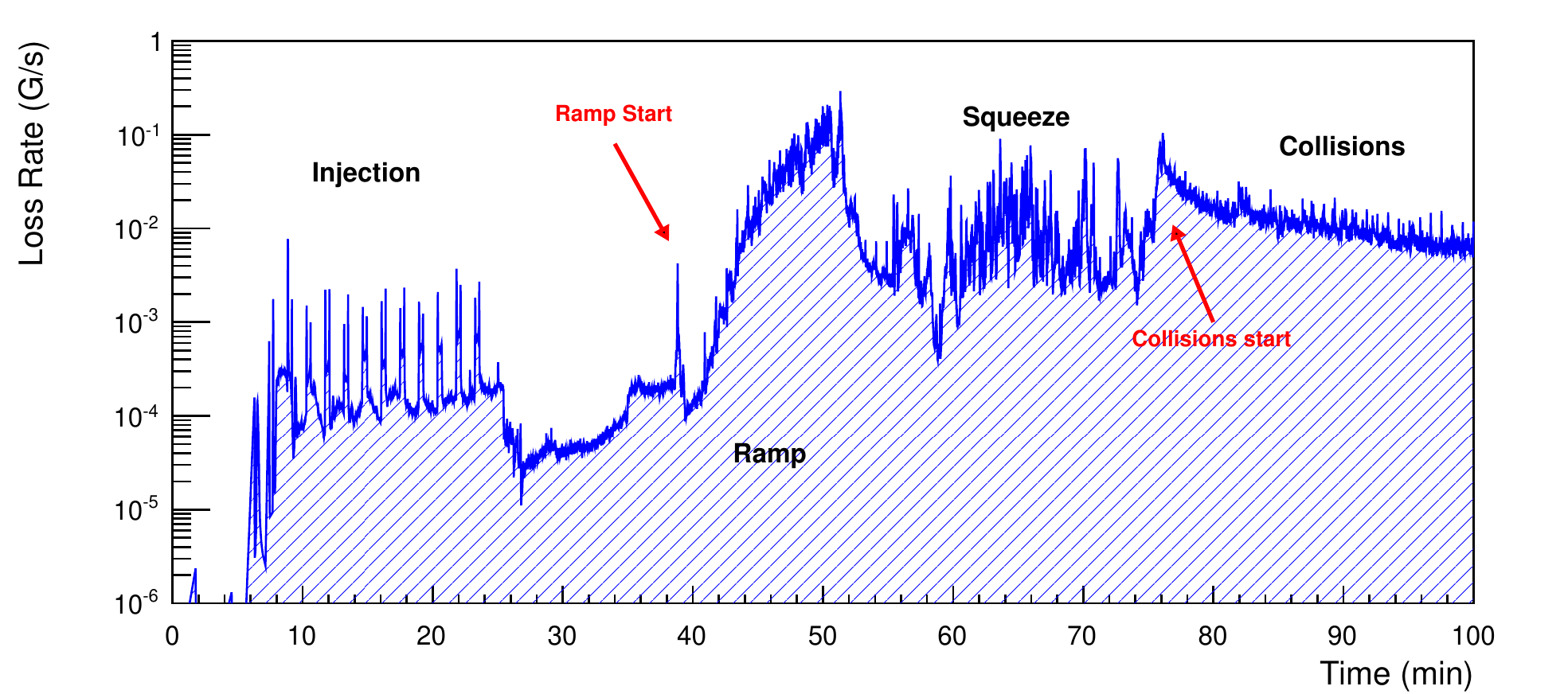}
   \end{center}
\caption{Beam loss rates averaged over 1~s at the primary collimator along a typical LHC cycle in 2012. During the squeeze phase the optics around the experiments is changed to reduce (squeeze) the betatron function at the IP.}
\label{fig:loss-cycle}
\end{figure}
The different types of losses are shown in Fig.~\ref{fig:loss-cycle} for a typical LHC cycle. The importance of the primary collimator opening appears clearly if the total beam intensity transmission presented in Fig.~\ref{fig:transmission} is compared between 2011 and 2012. In 2012 the collimators were set closer to the beam in order to protect a smaller aperture, which allowed smaller betatron functions to be reached at the IP and therefore smaller beam sizes at the collision points, yielding a 60\% higher luminosity~\cite{SQUEEZE}. The gap width has a strong impact on beam transmission and losses in the cycle.
\begin{itemize}
  \item In 2011 the intensity losses during the ramp and squeeze phases are so small that they cannot be measured with the current transformers. The losses are completely dominated by collisions. The primary collimator gaps correspond to an opening of $\pm 7.5$ $\sigma$, where $\sigma$ is the real beam size.
  \item In 2012, on the other hand, there are already significant losses during the ramp when the collimator gaps are closed to their high-energy setting. There are noticeable losses in the optics squeeze due to an increased sensitivity to even small orbit jitter at the level of 50~$\mu$m. The primary collimator gaps correspond to an opening of $\pm 5.2$ $\sigma$, where $\sigma$ is the real beam size.
\end{itemize}
The regular loss distribution during collisions is shown in Fig.~\ref{fig:loss-sb}. The beam loss rates near the experiments are almost as high as at the collimators. This is due to the fact that around the experiments the loss monitors record collision debris, the physics processes at very small angles that are not covered by the experiments. Their signals are proportional to the collider luminosity.

The thresholds of the LHC loss monitors are set to prevent quenches for monitors installed on superconducting elements and to prevent damage for room-temperature elements (for example collimators)~\cite{QUENCH}. In both cases some safety margin is desired.
The initial thresholds were set before LHC operation started based on rather coarse quench level estimates coupled to GEANT, FLUKA and MARS simulations~\cite{FLUKA}. During the first years of operation the thresholds were progressively adapted (many were increased) based on experience. Initially the thresholds on collimators were set to limit the average power loss significantly below the peak design power of 500~kW. The thresholds were only increased to match the nominal power loss once the operation of the LHC was well controlled and understood. Quench tests with wire scanners (nice point-like particle source), orbit bumps and short and high losses in the collimation area were used to determine more accurately the quench limits~\cite{QUENCH}.

\subsection{UFO losses}

\begin{figure}[btp]
  \begin{center}
\includegraphics[width=0.6\linewidth]{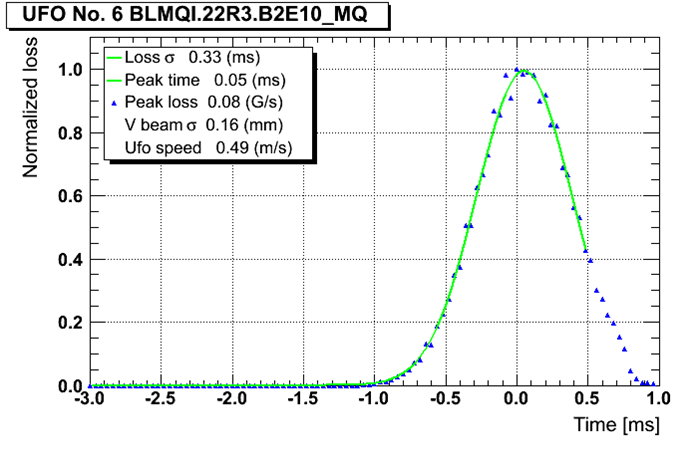}
   \end{center}
\caption{Time evolution of the beam loss during a UFO event. The vertical scale represents the beam loss relative to the highest loss during the UFO event.}
\label{fig:ufo-time}
\end{figure}

\begin{figure}[btp]
  \begin{center}
\includegraphics[width=1\linewidth]{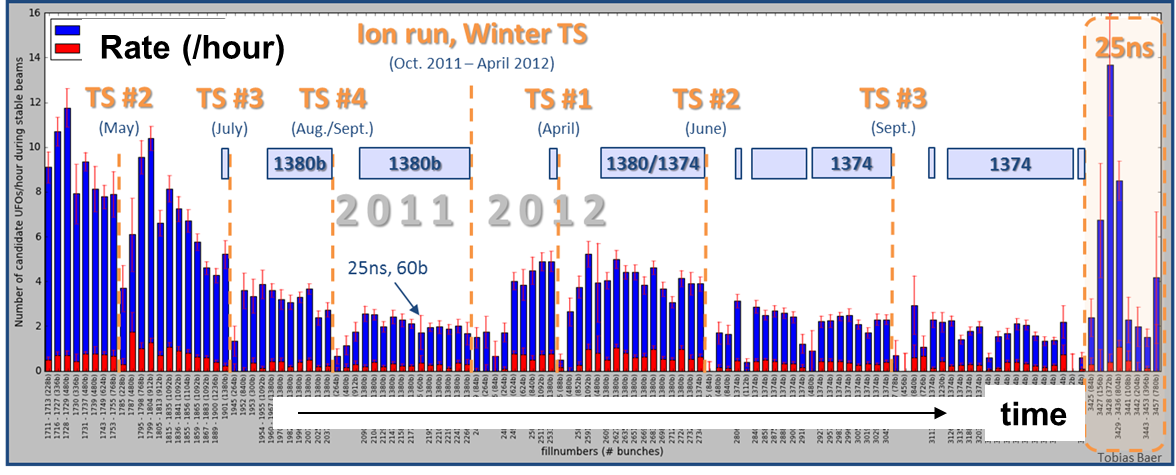}
   \end{center}
\caption{Evolution of the hourly rate of UFO events in 2011 and 2012. A slow conditioning is observed during each run (adapted from~\cite{UFOa,UFOb}).}
\label{fig:ufo-rate}
\end{figure}

Very fast beam loss events (time-scale of a millisecond and below; see Fig.~\ref{fig:ufo-time}) mainly in superconducting regions have been the \textit{surprise of LHC operation}. Those fast losses were nicknamed UFOs (unidentified falling objects) when it became clear that the UFO-type beam losses were due to small objects falling into the beam, the subsequent interaction of the beam particles leading to the showers and the beam losses~\cite{UFOa,UFOb}. The BLM signals are consistent with small (tens of $\mu$m diameter) dust particles `entering' the beam.
The vast majority of UFO events lead to losses below dump threshold, but around 20 beam dumps were triggered by such UFO-type events every year between 2010 and 2012. UFO events localized within the injection kickers could be traced to aluminium oxide dust present in large quantities on the inner surface of the ceramic vacuum chamber of those elements. A cleaning campaign of the kickers was made during the long shutdown in 2013--2014. UFOs occur mainly with high-intensity beams, but there is conditioning with beam; the event rate drops from 10 to 2 per hour over a year as shown in Fig.~\ref{fig:ufo-rate}. To improve the sensitivity of the LHC arc BLMs to UFO events, two out of six BLMs installed around the arc quadrupoles were re-located to the dipoles during the shutdown~\cite{BLM-CHAM15}. The monitoring and protection (quench prevention) capabilities of the BLMs are significantly improved with this re-location for the coming 6.5~TeV run. While the beam losses will increase due to the higher beam energy, the quench thresholds of the magnets will come down by a factor of 3--4, making the situation at 6.5~TeV more critical.

\section{Diagnostics and control}

The three main requirements for a modern MPS are:
\begin{itemize}
  \item protect the machine: the highest priority is obviously to avoid damage to the accelerator;
  \item protect the beam: complex protection systems reduce the availability of the accelerator; the number of `false' interlocks stopping operation must be minimized. This implies sometimes a trade-off between protection and operation;
  \item provide the evidence: clear (post-mortem) diagnostics must be provided when the protection systems stop operation or when
   something goes wrong (failure, damage, but also `near misses').
\end{itemize}

Once the MPS components have been commissioned, it is essential that the protection functionality is maintained during operation. Automated checks of the MPS components as pre- or post-flight checks can ensure that the MPS functionality is not degraded. For colliders with long cycle times there are two types of checks that fit well into the cycle, namely
pre-flight checks before injection and post-flight checks on data collected during a fill or during the beam dump (post-mortem data). Such tests can come in multiple forms, for example the verification of MPS-related settings such as interlock thresholds,
configuration checks of the beam interlock systems,
automated analysis of the faults and MPS reaction chain and
automated analysis of the dump system action. At the LHC the BLM system integrity checks must be performed at least once per 24 h period (or after the following beam dump if that interval is exceeded)~\cite{BLM-TEST}. The integrity checks are performed by a high-voltage modulation that is analysed to detect `dead' channels.

At the LHC the MPS is so critical that for every beam dump post-operation checks (POCs) are performed on the beam dump system post-mortem data (equipment and beam signals). Data collection and analysis are triggered automatically after each dump. The analysis covers  internal beam dump system signals and external beam information like dumped beam intensities, losses and beam positions in and around the dump channel. The analysis ensures that all signals are correct and that there is no loss of redundancy; the beam dumping system can then be considered `as good as new'. Machine operation is stopped if the beam dumping system POCs fail; an expert verification is required to re-start beam operation. This concept was so successful that it was extended to LHC injection in the form of automated checks of each injection quality.

Already during the design of the LHC MPS, post-mortem (PM) diagnostics was identified as a key component to understand the root causes of beam dumps~\cite{LHC-PM}. All key LHC systems implement circular PM data buffers that are frozen and read out when a beam dump is triggered.  The PM trigger is transmitted over the LHC machine timing system. The PM system also inserts an automatic entry in the LHC electronic logbook. Sampling frequencies of the data in the PM buffers range from ms or turn level to tens of milliseconds and are adapted to each system. Synchronization is critical to make sense of the data and define the event sequence; therefore time stamping is made at the source of the data. The LHC post-mortem data volume is currently around 200~MBytes. Since the analysis of the large volume of LHC PM data can be tedious, automatic analysis tools have been implemented to support operators and MPS experts in their work. The LHC PM server and graphical user interfaces are based on the JAVA language with standard interfaces to extract raw data and provide analysed data~\cite{LHC-PM-SERVER}.
This allows many persons to contribute to the analysis modules.
After a beam dump, injection into the LHC is blocked until the PM data is collected, pre-analysed (automatic) and signed by the operation crews. If an automated analysis identifies a critical problem, injection can only be released by a MPS expert. To detect any possible long-term trend MPS experts re-analyse all PM events collected above injection energy within a few days. This also provides a view that is independent of the operation teams.

\subsection{Settings control}

\begin{figure}[btp]
  \begin{center}
\includegraphics[width=0.9\linewidth]{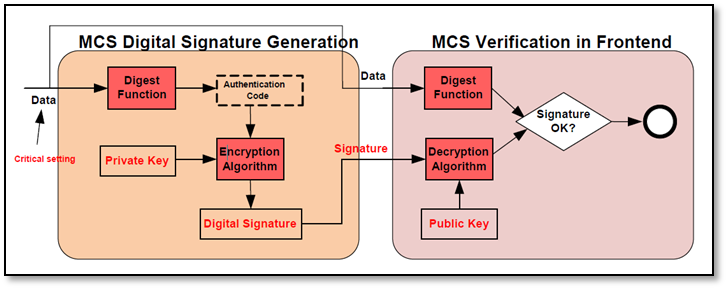}
   \end{center}
\caption{Principle of management of critical settings (MCS) with digital signature of the data}
\label{fig:mcs}
\end{figure}

Depending on its size, complexity and energy range an accelerator has a large volume of MPS settings in the form of BLM thresholds, current references and tolerances, etc. The management of settings associated to the MPS poses special problems, since access to the settings must be limited to authorized experts, yet it must be possible to inspect and download the settings from a central database to the front-end systems of the MPS. For the LHC access restrictions have been put in place based on \textit{role-based access control} (RBAC)~\cite{RBAC} and on digital signatures.
To protect MPS settings the concept of \textit{management of critical settings} (MCS)~\cite{MCS} was developed, a settings protection system that is fully embedded in the controls middleware and settings management. A setting that is defined as \textit{critical} has an associated digital signature. Only a user that has the appropriate RBAC role (MPS expert, BLM expert, etc) is able to generate the digital signature. The digital signature is generated at the moment when a setting is changed by an authenticated expert. The setting and the associated digital signature are transmitted together to the front-end computer: a critical setting is only accepted with a valid digital signature, see Fig.~\ref{fig:mcs}.

\subsection{Software interlocks}

The LHC MPS has inputs from many systems that operate independently and there is only very limited information exchange between the various systems. To implement interlocks on a global machine scale with correlation of data between many LHC systems, a software interlock system (SIS) was developed~\cite{SISa,SISb}. This system is able to collect information from any control device of the LHC and its injectors. The LHC SIS can dump one or both beams and inhibit LHC injection. It may perform global scale analysis among systems, correlate injector and LHC ring data for injection protection, etc.
New interlock tests can be implemented rapidly to protect against unexpected issues.
The LHC SIS is by design rather slow (reaction time of the level of seconds) but it is able to detect anomalies that could lead to problems in the future or prevent unnecessary beam dumps at injection. The SIS is based on a JAVA core server, with a large data buffer for currently up to around  2500 devices and settings. For each device data a timeout and no-data policy is defined.
Over 5000 tests are defined with the LHC SIS; tests can be defined as simple value comparisons or complex JAVA logic.

\begin{figure}[bth]
\begin{center}
\includegraphics[width=0.7\linewidth]{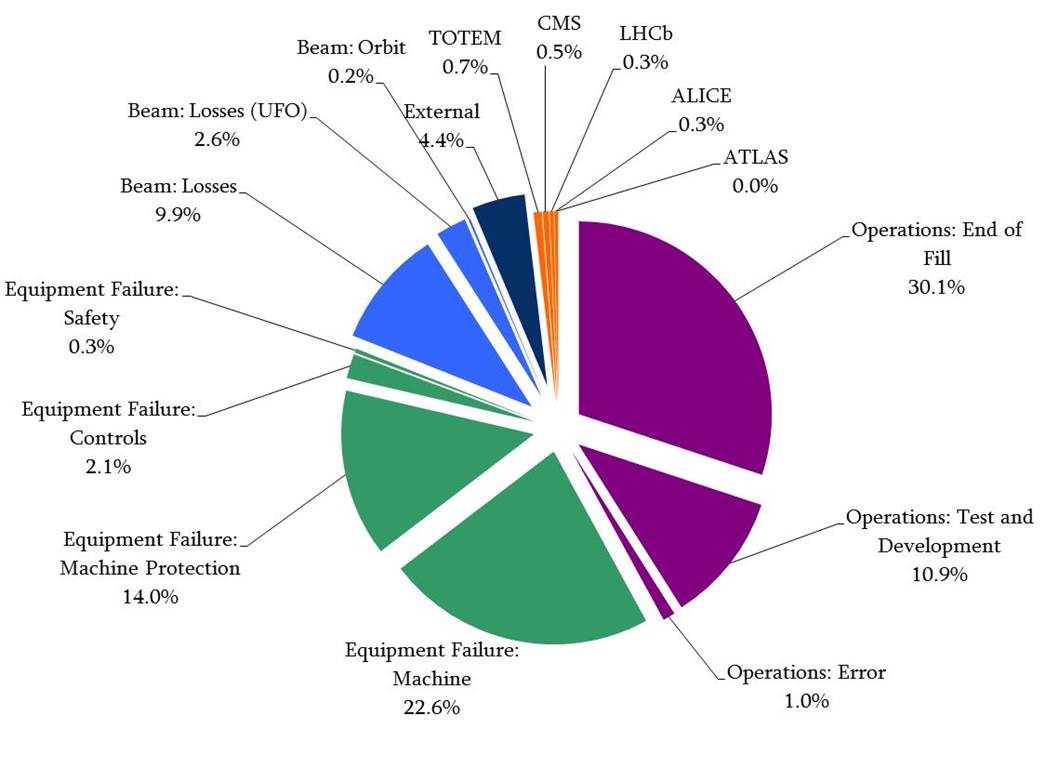}
\caption{Distribution of beam dump causes during the 2012 LHC proton run (adapted from~\cite{MPWS-2013}). The statistics include only beam dump above injection energy. }
\label{fig:dump2012}
\end{center}
\end{figure}

\section{Machine availability}

Peak performance in terms of luminosity is not the most important LHC parameter; the integrated performance is much more critical. To optimize the integrated time with collisions it is important to maximize the time spent colliding the beams and to minimize the time to re-establish collisions at the end of a fill (turn-around time). The fraction of time that the LHC spends colliding beams stably for the experiments amounted to 30--35\% of the total scheduled operation time~\cite{LHC-RUN1}. The average duration of collisions was only 6~h while the optimum duration would be in the range of 8 to 12~h. The reason why the fills are so short is because during high-intensity operation only one out of three fills is dumped by the operation crews; see Fig.~\ref{fig:dump2012}. The other two-thirds of the fills are dumped by the LHC MPS~\cite{MPWS-2013}. Roughly 14\% of the beam dumps are due to the failures of MPS subsystems (`false' dumps), with the following distribution of the MPS subsystems:
\begin{itemize}
  \item quench protection system (radiation to electronics): 65\%;
  \item BLM system:	13\%;
  \item beam dumping system: 12\%;
  \item software interlock system: 5\%;
  \item powering interlock system:	2.5\%;
  \item beam interlock system:	1.5\%.
\end{itemize}
A reliability working group predicted the rate of false dumps and the safety of the LHC MPS for 7~TeV operation before the LHC was switched on. This can now be compared with observations, but it must be observed that so far LHC operated at 4~TeV while the predictions were made for 7~TeV, which is not completely equivalent for certain systems. The observations are roughly in line with predictions, but some failures do not match completely; in particular, radiation to electronics was not included in the initial predictions~\cite{LBDS-AVAIL,LHCMP-AVAIL}.

\section{Machine experiments}

Machine experiments can be very exciting, but also risky periods for an accelerator. Frequently the machine is operated at some `distance' from standard conditions. For example collimator settings, orbit and optics may be changed. At the LHC every experiment is categorized according to the foreseen changes to the machine and to the beam intensity.
Experiments using intensities above the SBF limit must be prepared with a detailed written description of the changes to machines and the test procedures. In many cases the analysis of the document helped improve the efficiency of the experiment by spotting `impossible' things. This encourages experimenters to think about options with smaller MP footprint, for example lower intensity.

\section{Summary}

The LHC MPS has been very successful in protecting the LHC with over 100~MJ of stored beam energy while the LHC was operated at 4~TeV. No component was damaged by a failure leading to beam loss; the MPS therefore fulfilled its job. As expected, operation of the LHC was significantly constrained by MP due to the high stored energy and very low quench levels. Nevertheless, operation was rather smooth and the intensity ramp up from well below 1~MJ during the commissioning phase to over 100~MJ stored energy took only 2~weeks in 2012.
In 2015 the beam energy will be increased to 6.5~TeV: the stored energy of the beams will increase by a factor of 2 to 3, while the quench levels of the magnet drop by a factor of 3 to 5. This will be a new challenge for LHC operation, while at the same time the focus is shifting more and more towards high(er) machine availability.

\def\appendix#1#2{%
\setcounter{figure}{0}
\setcounter{table}{0}
\setcounter{equation}{0}
\def\thefigure{#1.\arabic{figure}}
\def\thetable{#1.\arabic{table}}
\def\theequation{#1\arabic{equation}}
\section*{Appendix\ #1:\ #2}%
}

\newpage
\appendix{A}{Powering incident in 2008}

Since the LHC powering incident in 2008 was the most severe damage that ever happened to an accelerator, a brief description of it is appended here even though this incident happened \textit{without beam}~\cite{PAC99}.

In the morning of 19 September 2008 the last commissioning step of the main
dipole circuit (154 magnets) of sector~34 (between IP3 and IP4 in Fig.~\ref{fig:layout}) was started, a ramp to
9.3~kA which corresponds to a beam energy of 5.5~TeV. During the
ramp an electrical fault developed in the powering busbar interconnection at a current of 8.7~kA. A resistive voltage appeared and increased to 1~V after less than 0.5~s, leading to the power converter trip.
The current started to decrease in the circuit and the energy
discharge switch opened, inserting dump resistors in the circuit. In this sequence of events, the quench
detection, power converter and energy discharge systems behaved as
expected. No resistive voltage
appeared on the dipoles of the circuit, individually equipped with
quench detectors and the quench of a magnet has been excluded
as initial event.

\begin{figure}[ptb]
\begin{center}
\includegraphics[width=0.5\linewidth]{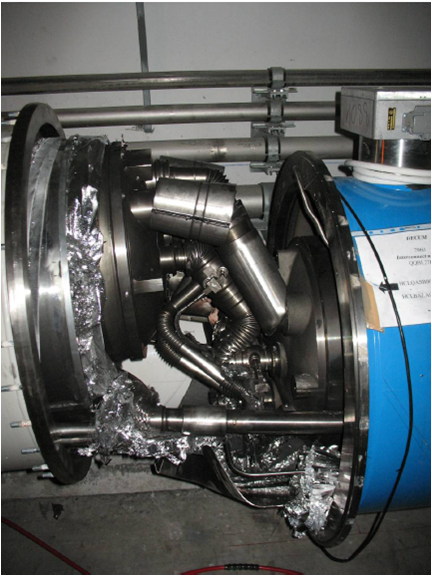}
\caption{A damaged interconnect between a quadrupole and a dipole magnet}
\label{fig:intercon}
\end{center}
\end{figure}

Within the first second, a main electrical arc and multiple smaller secondary arcs developed and punctured
the helium enclosure, leading to release of helium into the
insulation vacuum of the cryostat. After a few seconds the beam vacuum also degraded.
The spring-loaded
relief discs on the insulation vacuum enclosure opened when the pressure
exceeded atmospheric, thus relieving the helium to the tunnel. They
were however unable to maintain the pressure rise below the nominal
0.15~MPa absolute, thus resulting in large
pressure forces acting on the vacuum barriers separating neighbouring
subsectors (a subsector corresponds to two 107~m long cells), which damaged them as can be seen in
Fig.~\ref{fig:intercon}. The forces displaced dipoles in the
affected zone from their cold internal supports and knocked the
short straight section (SSS) cryostats housing the quadrupoles and
vacuum barriers from their external support jacks at three positions,
in some locations breaking the anchors in the
concrete floor of the tunnel. The displacement also damaged the connections to the
cryogenic distribution line. The main damage zone extends over approximately 700~m.

About 2~tons of helium were rapidly
released to the tunnel, producing a cloud which triggered oxygen
deficiency hazard detectors and
tripped an emergency stop, thus switching off all electrical power
 from sector~34. Before restoration of electrical power enabled the
actuation of cryogenic valves, another 4~tons of helium were lost at
lower flow rates. The total loss of inventory thus amounts to about
6~tons out of 15~tons initially in the sector.

\begin{figure}[pbh]
\begin{center}
\includegraphics[width=0.7\linewidth]{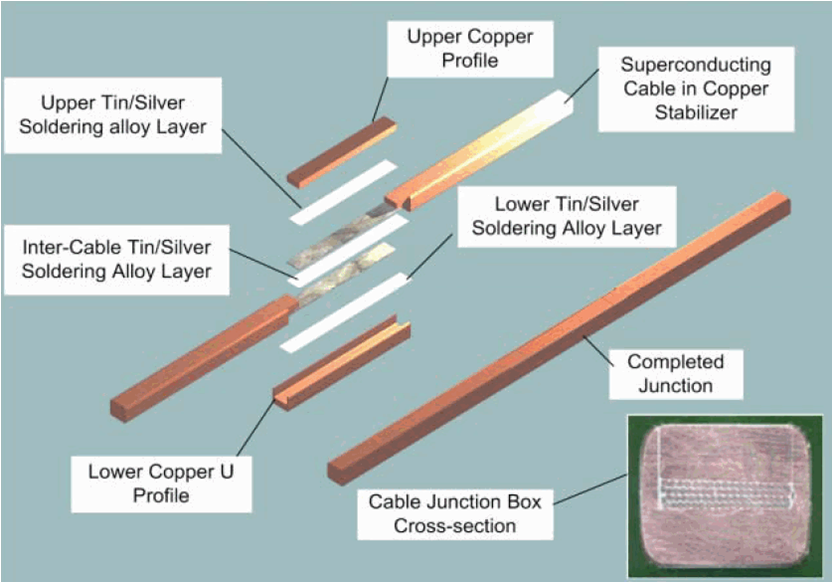}
\caption{Schematic of the main dipole busbar joint. The superconducting cable is
embedded in a copper stabilizer. At the joint the two busbar ends are inserted with
solder into a copper profile and welded.} \label{fig:joint}
\end{center}
\end{figure}

\begin{figure}[tbh]
\begin{center}
\includegraphics[width=0.35\linewidth,angle=-90]{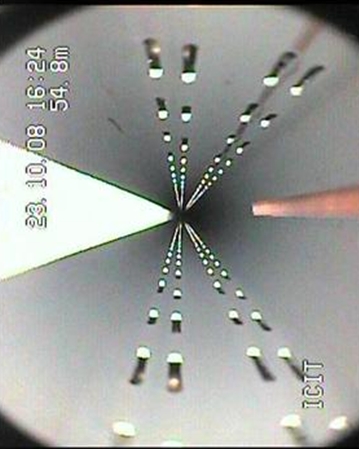}
\includegraphics[width=0.35\linewidth,angle=-90]{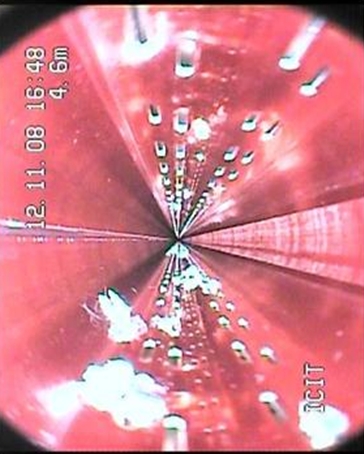}
\caption{Example of a vacuum chamber beam screen covered with soot
from the incident} \label{fig:vac-sooth}
\end{center}
\end{figure}

A post-mortem analysis of cryogenic temperature data
revealed a significant temperature anomaly in sector~34 during a
powering step to 7~kA performed a few days before the incident. A
steady temperature increase of up to 40~mK occurred in the
cryogenic cell of the incident. The excess power in the incident cell corresponds to an unaccounted resistance of
around $220~\mathrm{n}\Omega$. Given the location of the primary electrical arc,
the most likely hypothesis for the cause of the incident is a problem of the busbar joint.
The structure of such a joint is shown in Fig.~\ref{fig:joint}. The joints are brazed but not clamped and
the nominal joint resistance is $0.35~\mathrm{n}\Omega$. The incident could be reproduced in simulation
assuming a bad electrical and thermal contact of the copper stabilizer at the joint due to lack of solder
or poor-quality brazing~\cite{INCIDENT-REPORTS}.

A total of 53 magnets, 39~dipoles and 14~quadrupole SSSs had to be removed from
the tunnel and brought to the surface for cleaning and repair. Most of them were replaced with spare magnets.
All magnets have been thoroughly re-tested before re-installation in the tunnel.

\begin{figure}[btp]
  \begin{center}
\includegraphics[width=0.75\linewidth]{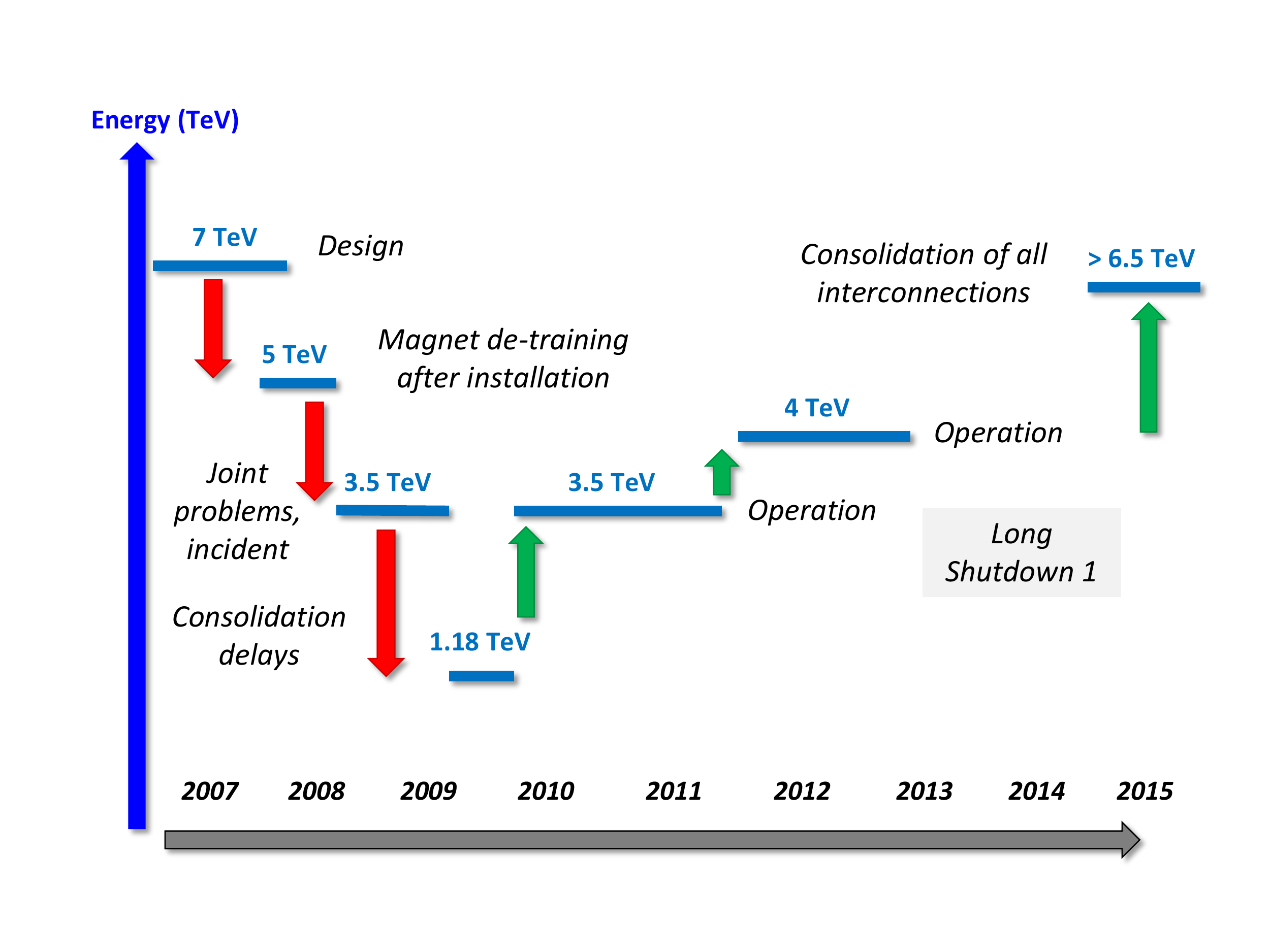}
   \end{center}
\caption{Evolution of the LHC beam energy from design to the present day}
\label{fig:energy-time}
\end{figure}

Both arc beam vacuum chambers were contaminated by soot from  electrical arcs and chips of multilayer
insulation over roughly 80\% of their length as can be seen in Fig.~\ref{fig:vac-sooth}.
Contamination by chips of multilayer insulation has been found over
long distances away from the position of the original incident.
These chips are deposited mostly on the beam screen surface, from where they are removed by in situ cleaning.

Following the incident and a review of the magnet and busbar protection system, a Quench Protection System (QPS) upgrade
was launched to protect all busbar joints of the arc main dipole and main quadrupole
circuits. The required voltage tabs were available, but a large volume
of electronics had to be developed. The busbar protection was finally operational in the spring of 2010 when the LHC was commissioned at 3.5~TeV. The evolution of the LHC beam energy over time is shown in Fig.~\ref{fig:energy-time}, in 2015 the energy will be pushed to 6.5~TeV after the consolidation campaign of the 2~year long shutdown.

\end{document}